\documentclass[aps,pre,reprint,superscriptaddress,nofootinbib]{revtex4-1}
\usepackage{graphicx}
\usepackage{amsmath}
\usepackage{amsfonts}
\usepackage{hyperref}
\newcommand{\Wi}{\textrm{Wi}}

\begin{document}

\title{A Force-Level Theory of the Rheology of Entangled Rod and Chain Polymer Liquids. I. Tube Deformation, Microscopic Yielding and the Nonlinear Elastic Limit }

\author{Kenenth S. Schweizer}
\email[]{kschweiz@illinois.edu}
\affiliation{Department of Materials Science and Department of Chemistry, University of Illinois, 1304 West Green Street, Urbana, IL 61801, USA}
\author{Daniel M. Sussman}
\affiliation{Department of Physics and Astronomy, University of Pennsylvania, 209 South 33rd Street, Philadelphia, Pennsylvania 19104, USA}

\date{\today}

\begin{abstract}
We employ a first principles, force-level approach to self-consistently construct the anharmonic tube confinement field for entangled fluids of rigid needles and for primitive-path (PP) level chains in two limiting situations where chain stretching is assumed to either completely relax or remain unrelaxed. The influence of shear and extensional deformation and polymer orientation is determined in a nonlinear elastic limit where dissipative relaxation processes are intentionally neglected. For needles and PP-level chains, a Gaussian analysis of transverse polymer dynamical fluctuations predicts that deformation-induced orientation leads to tube dilation. In contrast, for deformed polymers in which chain stretch does not relax we find tube compression. For all three systems, a finite maximum transverse entanglement force localizing the polymers in effective tubes is predicted. The conditions when this entanglement force can be overcome (a force imbalance) by an externally applied force associated with macroscopic deformation can be crisply defined in the nonlinear elastic limit, and the possibility of a ``microscopic absolute yielding'' event destroying the tube confinement can be analyzed. For needles and contour-relaxed PP chains, this force imbalance is found to occur at a stress of order the equilibrium shear modulus and thus a strain of order unity, corresponding to a mechanically fragile entanglement tube field. However, for unrelaxed stretched chains, tube compression stabilizes transverse polymer confinement, and there appears to be no force imbalance. These results collectively suggest that the crossover from elastic to irreversible viscous response requires chain retraction to initiate disentanglement. We qualitatively discuss comparisons with existing phenomenological models for nonlinear startup shear, step strain, and creep rheology experiments. 
\end{abstract}

\maketitle

\section{Introduction}
\subsection{Fundamental Open Questions and the Phenomenological State-of-the-Art}
The slow anisotropic dynamics and spectacular viscoelasticity of entangled macromolecular liquids of diverse architectures is a very difficult problem from a first-principles, many-body physics perspective [1,2]. These phenomena are widely believed to be a generic, dynamically emergent consequence of polymer connectivity and uncrossability on intermediate length and time scales, and the canonical view is that equilibrium structural correlations are unimportant at zeroth order. The leading phenomenological model for linear polymers is built on assumptions about ``topological entanglements'' and postulates for their consequences on polymer motion [1-5]. The seminal ansatz of Edwards and deGennes is that entanglements induce transverse confinement (mesoscopic dynamic localization) of a polymer in a tube (of diameter $~$ 3-10 nm in chain melts [6]). As a result, the liquid possesses a rubber-like entropic elasticity which persists on time scales shorter than the (long) time for the polymer to achieve Fickian diffusion via reptation, a 1-dimensional, anisotropic curvilinear Brownian motion along the coarse-grained polymer contour which is called the ``primitive path'' (PP) [4].

Traditional reptation-tube theory does not microscopically deduce nor construct the transverse confinement field that defines the tube from first principles, and the tube is not treated as an intrinsically dynamical object. It argues that on distances less than the tube diameter, polymers move isotropically and do not ``feel'' entanglements. Based on the static tube ansatz, scaling predictions can be made for the dependence of transport and relaxation properties on the polymer fluid density and the degree of polymerization; the key control parameter is the ratio of the tube diameter to the global polymer size [1-5]. Decades of development of this approach, with additional non-reptative mechanisms introduced by postulate, account well for a range of experimental observations under equilibrium conditions [5].

The situation for nonlinear rheology is far less settled, as additional assumptions must be made about what entanglements are and how they and individual polymers respond to strong deformations and polymer orientation. In the original Doi-Edwards (DE) theory [3], the tube responds in an affine manner, corresponding to a mechanically unbreakable confinement field. The most advanced phenomenological theory, the GLaMM (Graham-Likhtman-McLeish-Milner) model [7], builds on the DE approach and includes the postulated convective constraint release (CCR) effect of Marrucci [8], whereby strong-enough imposed shear flows weaken or destroy tube confinement. The CCR idea has been formulated in diverse ways [5,7-13]. It involves at least one adjustable parameter of a priori unclear magnitude, and \emph{qualitative} predictions for chain polymer rheology can be sensitive to its \emph{quantitative} value [14]. 

How valid the assumptions of the phenomenological models are in the nonlinear regime remains debated. Open questions include the following [5,15-25]. Do chains that have been stretched by deformation freely retract in a Rouse-like manner and on Rouse-like time scales? Can the transverse dynamical constraints soften or break under deformation-induced stresses? Are stress overshoots a signature of disentanglement and a crossover from elastic to viscous behavior, or are they due to an affine deformation of the tube? Directly answering such questions experimentally remains a large challenge, and incisive simulations to probe them have only very recently begun.

We believe that definitive theoretical progress requires a fundamental, ``bottom-up'' treatment of the entanglement problem on the tube diameter length scale. This has been our recent focus [26-33] Ð addressing in a self-consistent, first-principles, force-level manner how a confining tube dynamically emerges, what its properties are, and how it changes when polymers orient and/or are subjected to applied strain or stress. Specific questions considered include the following. What is the spatial nature of the tube confinement field in equilibrium? How is it modified by external forces, and can it be completely destroyed?  What is the fundamental physics underlying CCR in fast flows? What is the physical meaning of a stress overshoot? Such questions have been posed and indirectly probed experimentally by Wang and coworkers [17-22] for entangled chain polymer solutions and melts. They have proposed qualitatively new Ð and heatedly debated [14,23,25] Ð answers to some of these questions along with a new phenomenological model. 

\subsection{Microscopic Approaches}
The search for a deeper understanding has been recently pursued in equilibrium using simulation. Attempts to \emph{dynamically} identify what an entanglement is that go beyond the static primitive path (PP) idealization are notable. The idea that entanglements are rare, long-lived, primarily binary dynamic interactions between pairs of chains has been suggested [34-37]. Much less such work exists under conditions of strong deformation [38].

On the theoretical side, diverse attempts over the years have been made to analyze the equilibrium dynamics of entangled systems using time-dependent statistical mechanics. For instance, approaches based on a single-chain generalized Langevin equation (GLE) were proposed where the physics enters via an ad hoc viscoelastic memory function [39,40]. A first principles, force-level GLE approach using a polymer mode coupling theory idea [41,42] or renormalized Rouse methods [41-44] were extensively developed, with promising results. However, the assumptions common to these approaches Ð isotropic motion of chains, strong structure-dynamics connections, and a non-self-consistent calculation of the memory function matrix Ð expose them to significant criticisms; attempts to construct anisotropic GLE descriptions are highly phenomenological [45-47]. Self-consistent GLE-based ideas for the isotropic collective dynamics of a cluster of chains on the correlation-hole scale have been proposed which employ nonsingular effective potentials [48,49]. Most recently, a phenomenological theory based on analogy to the Higgs mechanism and Chern-Simon topological field theory has been proposed [50], though to date the concrete results of this program are limited. 

To our knowledge, none of the above approaches have addressed nonlinear rheology. Moreover, none self-consistently derive an anisotropic tube field at the level forces, and none exactly enforces dynamic uncrossability at any level. The exception is SzamelÕs equilibrium theory [51] for a fluid of non-rotating needles. This has been the starting point of our recent efforts for rigid stars and rods, and also flexible chains coarse-grained to the PP level. There have been 3 levels of development, which we summarize for context.

The first level was initiated by Szamel based on an ensemble-averaged dynamical approach [51]. The theory is self-consistently closed for single rod motion, and dynamic uncrossability at the two-rod level is \emph{exactly} included. Tube localization and long time reptation scaling laws emerge from the microscopic calculation. This approach, and its subsequent development by us, addresses the call of Arthur Lodge [52] for ``É.a self-consistent theory where the tube affects a chain, but motion of a chain affects the tube.''  It also can be viewed as a realization of the DE suggestion [2] (paraphrasing) that ``Éthe tube should be derived from basic equations by a kind of mean field approximation where it emerges as a dynamical, not static, concept, and thus is perhaps better understood as representing the effect of dynamical correlation of the environment rather than the usual mean field.'' The Gaussian closure of SzamelÕs theory implies a harmonic confinement field, and thus large amplitude transverse polymer fluctuations are not captured.

In the second level, we addressed the problem of non-Gaussian transverse confinement based on a distinct stochastic trajectory [26-28] approach built on nonlinear Langevin equation (NLE) ideas [53] originally formulated for glassy dynamics. It is self-consistent in the sense that the tube controls how a tagged polymer anisotropically moves in space, but the motion of polymers modifies the tube constraints. As a consequence, the confinement field is predicted to be very anharmonic, weakening as polymers displace in the transverse direction. This effect has qualitatively novel consequences; for example, the entanglement network is characterized by a maximum restoring force and thus the transverse dynamical constraints can be overcome by external forces [27,31]. Predictions for this anharmonic aspect were successfully tested against equilibrium experiments on solutions of the rod-like biopolymer F-actin [54] and against simulations of entangled chain melts [36] analyzed at the PP level. Our ideas have also been invoked to qualitatively interpret recent experiments on entangled DNA solutions [55]. 

In the third level of development we generalized our approach to treat the nonlinear rheology of entangled rod solutions subjected to step strain [32] and continuous startup shear [33] deformations, again guided by advances in glass physics [53,56]. The tube confinement field is nonlinearly coupled to polymer orientation, polymer (reptative) motion, and the macroscopic stress-strain response. The new concept of ``absolute microscopic yielding'' [31-33] was proposed, corresponding to the destruction of the entanglement network by an applied force. This leads to qualitatively new physical behavior not captured by phenomenological tube models.

To date our nonlinear rheological work has only been for rigid rod solutions, the simplest entangled systems which retain the generic mechanism of stress storage via polymer orientation. A virtue of rods is that chain stretch and retraction are irrelevant, and attention can be focused on the purely ``transverse'' aspect of entanglement physics. Although entangled rod solutions have received far less theoretical, simulation and experimental attention than flexible chain liquids, they are very important in diverse biopolymer contexts, and also germane to liquid-crystal-forming synthetic polymers. Moreover, we believe they are relevant to chains in relatively slow flows. This was the viewpoint of Doi and Edwards [2,3], who stated that their seminal work had more in common with rods than coils given their assumption of rapid contour length equilibration and adoption of a rigid PP description. Moreover, DE introduced the ``independent alignment approximation'' (IAA) [2,3] which allows the behavior of rigid PP steps to be related to the full rheological response of chain melts. However, DE theory fails to predict a physical flow curve (it predicts excessive shear thinning), which was the original motivation for the CCR proposal. Although DE theory also predicts an unphysical flow curve for entangled rods, no formulation of CCR for rod fluids has been proposed. Advanced phenomenological theories (such as GLaMM [7]) relate CCR to chain retraction, so the mechanism for introducing the CCR concept in classic tube models of rod solutions is unclear.

For the above reasons, we argue that fluids of rigid rods are the most fruitful starting point in an initial search for a fundamental, force-level theory of entangled polymer rheology. Per EinsteinÕs mantra, we are motivated by the view that ``everything should be as simple as possible, but not simpler.'' Our aim is not to a priori argue whether various ansatzes of phenomenological tube models for nonlinear rheology are false, but rather to use a first-principles approach to explore the validity of existing conjectures.

\subsection{Goals and Outline}
The present article and its companion paper are the first two of a planned four-part series aimed at developing a predictive microscopic approach to the nonlinear rheology of entangled rod and chain polymer liquids in slow and fast flows. We focus initially on the heavily entangled limit. Given how technically unconventional our approach is for the polymer community, one goal of the first two papers is to expose the physical content of our approach more intuitively and with greater clarity than presented so far. Thus, relevant prior results are first reviewed. New analytic analysis and numerical calculations are then presented for entangled rod solutions and for chains on time scales both long and short relative to contour-length relaxation. This first article (paper I) focuses on how orientation and chain deformation modifies the tube confinement field, and how such effects, in concert with the direct consequences of external forces, can lead to ``microscopic absolute yielding.'' The latter analysis is carried out in a ``nonlinear elastic network'' limit where irreversible relaxation processes are neglected.  

Section II reviews our key ideas for entangled rods in equilibrium, emphasizing the conceptual content and analytic results. In section III, the special limit we call the nonlinear elastic regime is introduced and analyzed. The concept of microscopic yielding as a force imbalance on the tube length scale can be precisely defined, in analogy with granular (and some glassy) materials. Its connection to fast startup continuous shear experiments, and to creep and step-strain measurements, is explained. The extension of the approach to treat tube deformation and microscopic absolute yielding of chain liquids within the PP framework is studied in sections V and VI, which, respectively, consider the limits where chain stretch relaxes rapidly or slowly on the relevant time scales. The paper concludes with a brief discussion in section VII. In the following paper (paper II) we address the effect of nonlinear deformation and flow on reptation and the tube survival function, formulate a full dynamical treatment of startup continuous shear rheology, and analyze steady state tube dilation effects in polymer melts. 

Throughout this article we do not repeat the well-documented statistical mechanical derivations previously presented [29, 51]. Rather, we focus on the underlying physical assumptions of our approach and their implications, consistent with our aim that the papers be useful for theorists, simulators, and experimentalists. It also will serve as the conceptual foundation for formulating a full theory of the nonlinear rheology of stretchable chain liquids in slow and fast flows in subsequent papers III and IV.

\section{Background: Entangled Dynamics in Equilibrium}
\subsection{Minimalist Model}
We consider the minimalist model of a gas of infinitely thin, rigid, dynamically uncrossable, non-rotating needles of length $L$ and dimensionless number density $\rho {{L}^{3}}$. The space-filling volume fraction is zero, and hence there are no equilibrium spatial correlations. The focus is on the heavily entangled limit, far beyond the entanglement crossover (at ${{\rho }_{e}}$), $\rho {{L}^{3}}\gg{{\rho }_{e}}{{L}^{3}}\approx 10-20.$ The needles are dissolved in an implicit solvent, and in the absence of mutual interactions they undergo Fickian diffusion governed by dilute solution hydrodynamics, ${{D}_{\bot}}={{D}_{0,\bot}}$, ${{D}_{||}}={{D}_{0,||}}$. Rotations are neglected for technical reasons, but in the heavily entangled limit we believe this does not introduce qualitative errors since the slow perpendicular CM and rotational motions are slaved [2]. However, by ignoring rod rotation we expect this model to quantitatively over-predict the dynamic consequences of needle uncrossability.  

\subsection{Equilibrium Theory: Gaussian Level}
The ensemble-averaged dynamics are characterized by an infinite hierarchy of distribution functions obeying coupled Smoluchowski evolution equations [51]. The uncrossability constraint on trajectories enters via (collisional) T-operators which play the role of forces. The resulting equilibrium dynamics problem, and the theoretical simplifications adopted to render it tractable, are sketched in Figure 1. Diffusing needles have stochastic trajectories interrupted and changed via instantaneous binary collisions. At high densities, these collisions become strongly correlated in space and time, and tube localization and anisotropic reptation dynamically emerges (i.e., is not a priori assumed) from these correlated collisions. 

\begin{figure}
\centerline{\includegraphics[width=0.9\linewidth]{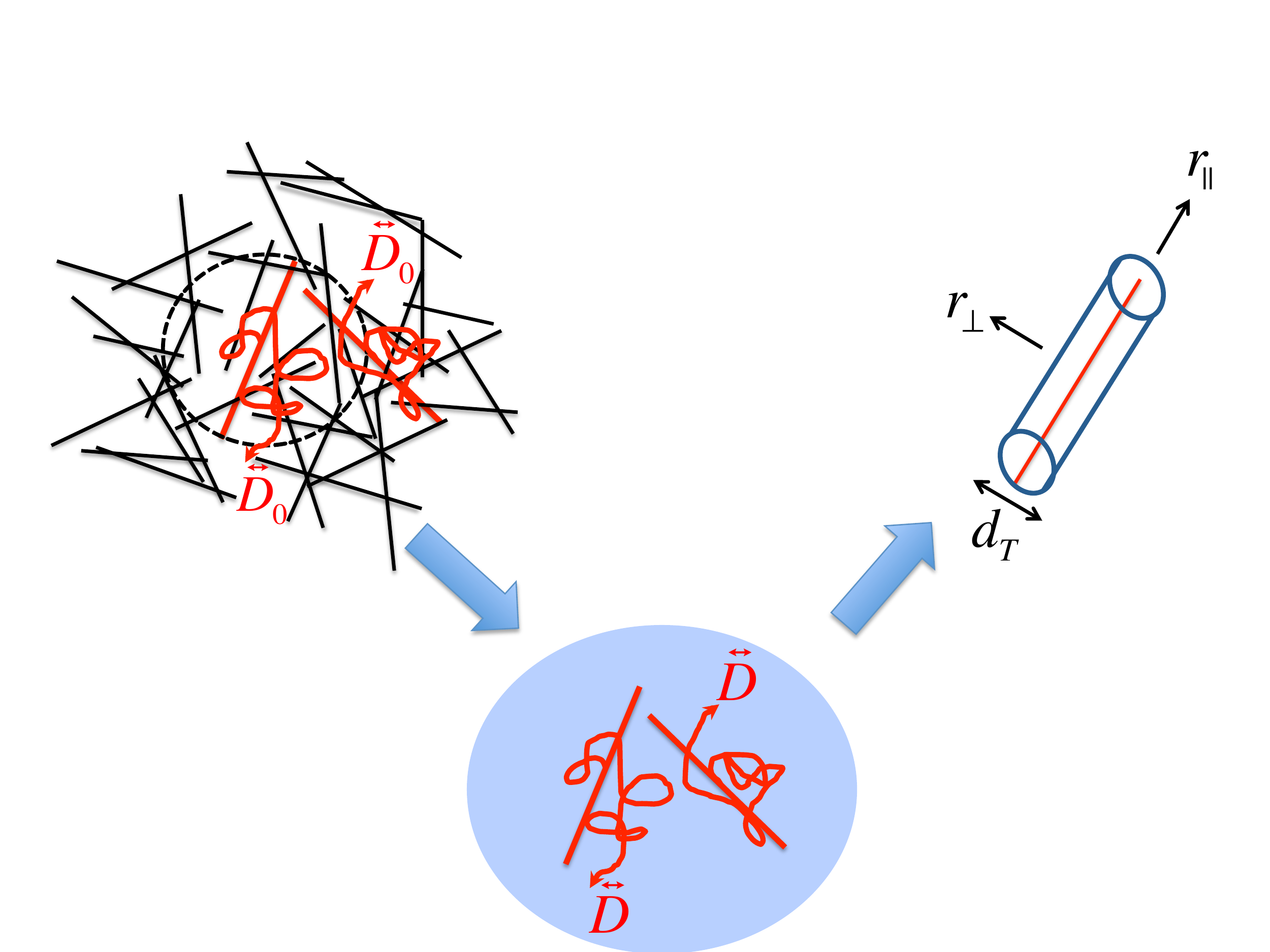}}
\caption{Conceptual schematic. (a) The initial model system studied consists of infinitely thin, uncrossable, rigid, non-rotating needles in solution (implicit solvent). In the absence of interactions the needles execute Fickian diffusion with a dilute solution diffusivity $D_0$. Collisions (dynamic uncrossability) correlate needle motions on a spatial length scale of order their length, as indicated by the dashed circle. (b) Binary collisions are exactly included in the theory and all higher order terms are approximately treated, corresponding to a self-consistent dynamical mean-field picture at the two-rod level. The diffusivity which controls transport is renormalized, $D_0\rightarrow D$, where $D$ is the full, to be self-consistently determined, long time self-diffusivity. (c) At the single-needle level a tube confinement field emerges that controls anisotropic CM needle displacements.}
\end{figure}

 	To treat the space-time correlation of arbitrary sequences of collisions involving many rods, a one-rod self-consistent dynamic mean field closure is adopted [51]. The dilute solution diffusion constants that control the rate of encounter of a pair of isolated needles is replaced with an effective diffusion tensor, $\overset{\lower0.5em\hbox{$\smash{\scriptscriptstyle\leftrightarrow}$}}{D}(z)$, where $z$ is a Laplace transform variable conjugate to time. This approximately captures the higher order collision processes. It is of the dynamic Gaussian form since $\overset{\lower0.5em\hbox{$\smash{\scriptscriptstyle\leftrightarrow}$}}{D}(z)$ is directly related to a generalized second moment of the needle center-of-mass (CM) mean square displacement. The inverse of $\overset{\lower0.5em\hbox{$\smash{\scriptscriptstyle\leftrightarrow}$}}{D}(z)$ is a generalized friction constant which can capture, in the appropriate time regimes, both transient harmonic tube confinement and long time diffusion.  

The technical implementation of the above ideas results in a self-consistent equation for $\overset{\lower0.5em\hbox{$\smash{\scriptscriptstyle\leftrightarrow}$}}{D}(z)$ involving an integral over the structural correlations of the needle fluid. Since the needles are infinitely thin, their CM have ideal gas structural statistics. We do allow for the possibility [27-29], however, that the needles have an (externally) imposed orientational order, so that their equilibrium pair correlation function with orientational distributions $\alpha ({{\vec{u}}_{j}})$ (over unit vectors ${{\vec{u}}_{j}}$) is in general ${{g}_{2}}(\vec{r},{{\vec{u}}_{1}},{{\vec{u}}_{2}})=\alpha ({{\vec{u}}_{1}})\alpha ({{\vec{u}}_{2}})$. We emphasize four key points. (i) The equilibrium CM spatial structure of the fluid is random; in particular, there is no long-range deGennes correlation hole [1]. (ii) The dynamical consequences of isolated binary collisions on polymer trajectories are exactly included. (iii) There is a self-consistent feedback between the diffusion of a tagged polymer and the forces it experiences due to the other moving polymers. (iv) The effective transverse diffusivity, ${{D}_{\bot }}$, is coupled to the parallel diffusion constant, with the result that CM motion is ultimately controlled by a reptative process. 

The resulting self-consistent equation for $\overset{\lower0.5em\hbox{$\smash{\scriptscriptstyle\leftrightarrow}$}}{D}(z)$ can be numerically solved, and analytic results can be derived in certain situations. The simplest analytically tractable case concerns a hypothetical fluid in which the needles diffuse isotropically in 3 dimensions. The long time isotropic diffusion constant in this scenario is [27]:
\begin{equation}
 D/{{D}_{0}}=1-(\rho /{{\rho }_{c,iso}}),\ \ \        {{\rho }_{c,iso}}{{L}^{3}}\approx 18.2
\end{equation}
Thus, for high-enough needle densities, isotropic motion is quenched, corresponding to a ``topologicalÕÕ glass transition. The latter is predicted to occur at a reasonable needle number density, corresponding to $(18.2/{{L}^{3}})(\pi {{L}^{3}}/6)\approx 10$ interpenetrating rods in a region of space of diameter $L$ [27,29]. Physically, this calculation signals a dynamic crossover: ``entanglementsÕÕ emerge and isotropic motion is quenched, heralding the onset of anisotropic diffusion.  

The second analytically solvable situation is for long time \emph{anisotropic} diffusion where the unconstrained parallel reptative motion (${{D}_{||}}={{D}_{||,0}}$) restores global ergodicity. One can self-consistently compute ${{D}_{\bot }}$, and a full numerical analysis from the low density (independent binary collision) regime to the high density (heavily entangled) regime yields an apparent dynamical crossover at [27,51] ${{\rho }_{e}}{{L}^{3}}\approx 10.1$. In the DE picture, the underlying physics controlling long-time transverse CM and rotational diffusion is qualitatively the same Ð each proceeds via a small lateral displacement or rotation when a rod reptates out of its local tube and then becomes trapped in another tube [2]. In the heavily entangled limit, the asymptotic scaling predictions of reptation are recovered:
\begin{equation}
{{D}_{\bot }}/{{D}_{0}}\propto {{D}_{rot}}/{{D}_{0}}\propto {{(\rho {{L}^{3}})}^{-2}}\propto {{\left( {{d}_{T}}/L \right)}^{2}},\quad \rho \gg{{\rho }_{e}}
\end{equation}
Here the connection to the tube diameter is a classic reptation-tube model result [2]. Our numerical calculations of the perpendicular diffusion constant are in excellent quantitative agreement with simulations after empirically rescaling $\rho $ by a crossover value 2-3 times larger than a priori predicted [27]. The direction of this difference is consistent with our neglect of rotations. If reptation is quenched, ${{D}_{||,0}}=0$, this calculation predicts a slightly different critical rod number density above which the system dynamically freezes [27]:
\begin{equation}
{{D}_{\bot }}/{{D}_{0}}=1-(\rho /\rho _{c}^{*}),\ \      \textrm{with}\ \  \rho _{c}^{*}{{L}^{3}}\simeq 9.3
\end{equation}

A third analytical calculation corresponds to \emph{a priori} quenching reptation, ${{D}_{\parallel ,0}}\to {{0}^{+}}$, and studying the system on intermediate time scales. This allows one to precisely define and self-consistently compute an emergent transverse localization length, ${{r}_{l}}\equiv {{d}_{T}}/2$, at the level of harmonic fluctuations [27,51]. A localized solution first occurs at ${{\rho }_{c}}{{L}^{3}}=3\sqrt{2}$, a factor of $\sim 2$ smaller than estimates based on the above isotropic or reptation-quenched transverse diffusivity. Such a numerical difference is unsurprising given the theory is approximate, and it seems consistent with experiments [1-6] which find that the value of the entanglement crossover depends on the measured property. We believe it is significant that all of the above theoretical calculations for the emergence of entanglement effects correspond to when $\sim 5-10$ rods interpenetrate. Far beyond threshold, the tube diameter has the qualitative DE scaling with density:
\begin{equation}
{{d}_{T}}=\frac{16\sqrt{2}}{\pi \rho {{L}^{2}}}=\frac{16\sqrt{2}}{9\pi }\frac{{{\xi }^{2}}}{L}<<\xi			
\end{equation}
where the geometric mesh size is [57,58] :
\begin{equation}
\xi =\sqrt{\frac{3}{\rho L}}
\end{equation}
The tube diameter is much smaller than the geometric mesh. This contrasts with both chain polymer liquids $(d_T\gg \xi )$ [2,4] and semiflexible systems like F-actin $(d_T\sim \xi )$ [57,58].

Finally, we can treat a gas of needles with an imposed orientational order parameter $S=\langle (3{{\cos }^{2}}\theta -1)/2\rangle$ per a nematic $(S>0)$ or discotic $(S<0)$ liquid crystal. Physically, alignment reduces the probability of rod collisions, and thus is expected to decrease the effective entanglement density. Using a simple Onsager-like distribution for the relative orientation of two rods, we predict for $S>0$ [27,29]:  
\begin{equation}
{{d}_{T}}(S)\approx \frac{16\sqrt{2}}{\pi \rho {{L}^{2}}}\cdot \frac{1}{\sqrt{1-S}}
\end{equation}
To a reasonable approximation, rod alignment simply modifies $\rho \to \rho \sqrt{1-S}$ in the isotropic theory, as long as $S$ is not too close to unity. A caveat is that a 3-d tensorial description is in principle relevant, e.g., the direction parallel to the rod axis is not in general the direction of uniaxial order; this complication has been averaged over in our calculation, as also done in simulation studies [30]. A corollary is that the perpendicular and rotational relaxation time are predicted to decrease as rods orient as [27,29]: 
\begin{equation}
\frac{{{\tau }_{rot}}(\rho ,S)}{{{\tau }_{rot}}(\rho ,0)}\approx \frac{{{\tau }_{\bot }}(\rho ,S)}{{{\tau }_{\bot }}(\rho ,0)}\equiv \frac{{{D}_{\bot }}(\rho ,0)}{{{D}_{\bot }}(\rho ,S)}={{\left( \frac{{{d}_{T}}(\rho ,0)}{{{d}_{T}}(\rho ,S)} \right)}^{2}}\approx 1-S
\end{equation}
where ${{\tau }_{rot}}(\rho ,0)\propto {{L}^{2}}/{{D}_{\bot }}(\rho ,0).$

All the above results are fully consistent with the reptation-tube model for isotropic rod liquids [2], now microscopically derived. We believe this provides a strong foundation for extending the approach to address anharmonic aspects of tube confinement in equilibrium and nonlinear rheological phenomena.  

\subsection{Nongaussian Effects: Nonlinear Langevin Equation Extension}
The Gaussian theory of the previous section effectively assumes an infinitely strong harmonic confinement field. This is akin to the classic unbreakable tube model ansatz [1,2], and more advanced phenomenological descriptions such as GLaMM [7], slip spring [59], and slip link [60] models. Is this assumption correct? Answering this requires a theoretical approach that can treat non-Gaussian transverse polymer displacements. We have addressed this based on the nonlinear Langevin equation (NLE) approach, which extends the Gaussian description to treat large amplitude trajectories. 
One begins by rewriting the self-consistent equation for the transverse localization, derived using the Gaussian approximation, as a transverse force balance:

\begin{equation}
{{f}_{\bot }}({{r}_{l}})=\frac{2{{k}_{B}}T}{{{r}_{l}}}-2{{r}_{l}}{{K}_{\bot }}({{r}_{l}})=0,
\end{equation}
\begin{equation}
\beta {{K}_{\bot }}({{r}_{l}})=\frac{\pi \rho L}{8\sqrt{2}}F\left( \frac{L}{{{r}_{l}}} \right)
\end{equation}
where $F(x)$ is a combination of Bessel and Struve functions [27,29]. The two terms on the right hand side of Eq. (8) quantify the delocalizing effects that favor unbounded transverse diffusion and the localizing force due to dynamic uncrossability (entanglements).  The effect of the latter dynamically enters as a nonlinear restoring force, where ${{K}_{\bot }}({{r}_{l}})$ is a monotonically decreasing function. The key physical idea is to invoke a Òlocal equilibriumÓ picture where the ensemble averaged localization length is replaced by the dynamic displacement variable, ${{r}_{\bot }}$. Thus, Eq. (8) is interpreted as a force balance at the instantaneous tagged-particle level. Predicting the transverse trajectory of a tagged rod requires additional noise and short-time friction terms, and in the overdamped limit this leads to [27,29]:
\begin{equation}
{{\zeta }_{0}}\frac{d{{r}_{\bot }}}{dt}=-\frac{\partial }{\partial {{r}_{\bot }}}{{F}_{dyn}}({{r}_{\bot }})+\delta {{f}_{s}}
\end{equation}
\begin{equation*}
{{F}_{dyn}}({{r}_{\bot }})=-\int_{{{r}_{l}}}^{{{r}_{\bot }}}{{{f}_{\bot }}(r)dr},
\end{equation*} 
where ${{\zeta }_{0}}$ is the short-time friction constant (solvent friction for rods), $\delta {{f}_{s}}$ the corresponding white noise random force, and ${{F}_{dyn}}({{r}_{\bot }})$ is a Òdynamic free energyÓ which corresponds to the full transverse tube confinement field due to rod dynamic uncrossability. By construction, in the absence of noise the NLE reduces to Eq. (8) in the long time limit. But, in general, polymer motion beyond the small-displacement regime self-consistently leads to a weakening of the tube confinement field. 

Figure 2 presents numerical results for a heavily entangled needle fluid of $\rho /{{\rho }_{c}}=1000$ that indicates the two key length scales (the transverse localization length, rl, and the location of the maximum restoring force, $r_m$) and the key force scale of the dynamic free energy. The confinement potential grows logarithmically for very large transverse displacements [27], corresponding to an infinite entropic barrier height. Although this agrees intuitively with the tube model picture, the precise asymptotic form is sensitive to technical assumptions. In practice the asymptotic behavior has little or no effect on our dynamical predictions in equilibrium [32,33]. Although the barrier grows without bound, we predict that the maximum entanglement restoring force is finite. This idea is not in existing phenomenological tube models, but it seems physically inevitable to us given that entanglements are not permanent chemical crosslinks as in a rubber. We believe that the relevant question, which is \emph{quantitative}, concerns the precise mechanical strength of the tube. One generically expects the answer is very important for the question of whether the tube is breakable (or massively weakened) in the nonlinear deformation regime. In the heavily entangled limit we find the maximum restoring force keeping a rod polymer in the tube is [27-29]:
\begin{equation}
{{f}_{max}}\approx 4\frac{{{k}_{B}}T}{{{d}_{T}}}
\end{equation}
Thus, the tube diameter determines its mechanical strength. If an external force of this magnitude or greater is applied along the transverse direction, then the confinement potential becomes a monotonically decreasing function of transverse displacement, signaling the destruction of the entanglement network. The confinement force reaches a maximum value at a transverse rod displacement that is much larger than the tube diameter, but which is comparable to the physical mesh size [27]:
\begin{equation}
r_{m}^{{}}\simeq \sqrt{\frac{4\sqrt{2}}{\rho L}}=\sqrt{\frac{4\sqrt{2}}{3}}\xi \ \ \gg{{r}_{l}}={{d}_{T}}/2
\end{equation}

\begin{figure}
\centerline{\includegraphics[width=0.9\linewidth]{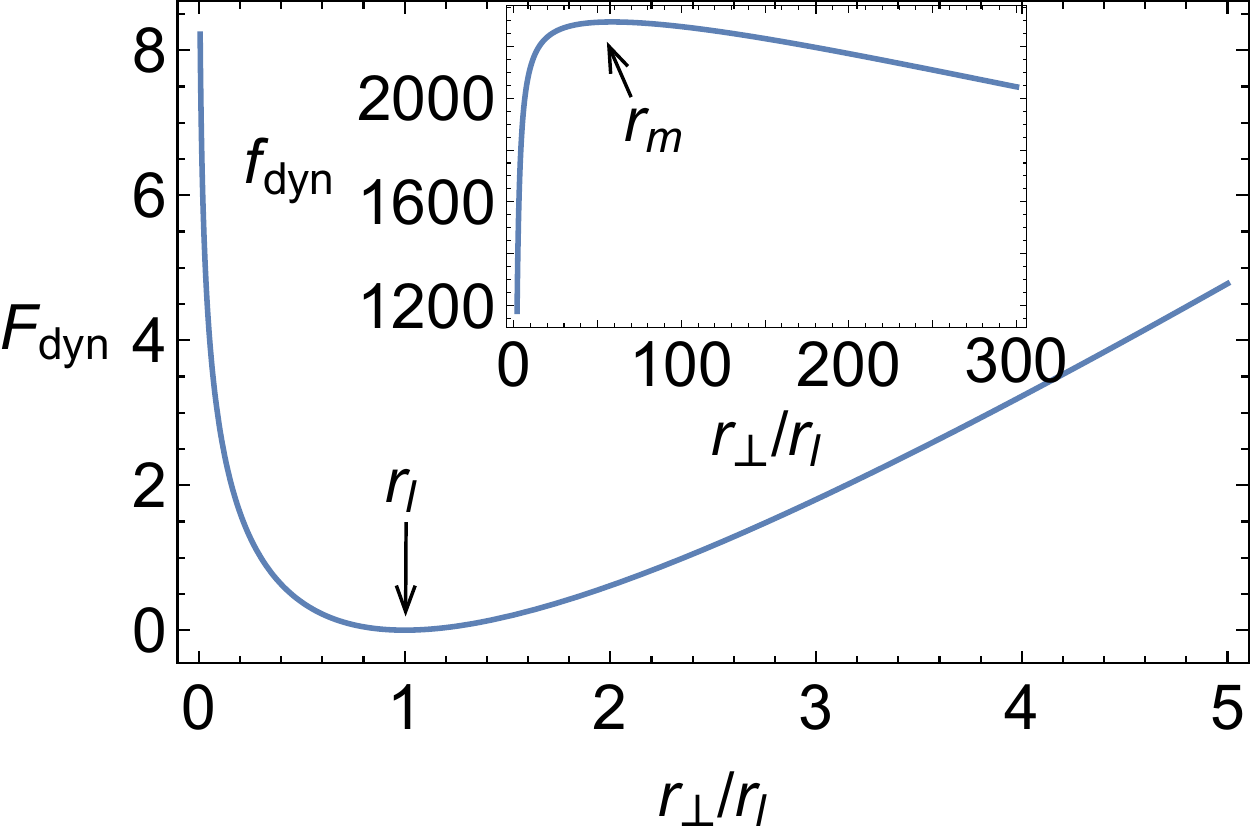}}
\caption{Anharmonic dynamic free energy (transverse tube confinement field) in units of the thermal energy versus transverse needle center-of-mass displacement normalized by the displacement corresponding to the minimum of the curve, for a very heavily entangled system with $\rho/\rho_c=1000$ . At this dimensionless density the tube diameter is $d_T=2r_l\approx0.0017L$ .  Inset: Effective transverse force localizing a needle in the tube in units of thermal energy per needle length. A maximum value occurs at a displacement $r_m$  that is much greater than the minimum (tube radius) but much less than the needle length. }
\end{figure}

The full anharmonic confinement field defines the distribution of transverse displacements at intermediate times (when polymers are localized in their tubes) via $P({{r}_{\bot }})\propto \exp \left( -\beta {{F}_{dyn}}({{r}_{\bot }}) \right)$. Predictions for this quantity are in quantitative agreement with experiments on F-actin solutions [54], where the observed specific form of the non-Gaussian distribution is consistent with the idea of a finite maximum mean tube restoring force. This quantitative agreement is especially significant since finite tube strength plays an essential role in our non-traditional theory of rheology. Extension of the rod theory to the PP chain level [28] leads to results for the full PP step-length distribution in good agreement with simulations [36] and qualitatively consistent with DNA experiments [55].

\section{Nonlinear Elastic Limit}
The most basic question in our approach to nonlinear rheology is how a force applied to all polymers in the material modifies the tube confinement potential. Here we recall the general formulation, and then apply it in a limiting situation where relaxation processes are absent. In this Ònonlinear elasticÓ limit we treat a heavily entangled material as a physical network, and analytic predictions are possible. This analysis is experimentally relevant in 3 limiting situations. (i) The large dimensionless rate (Weissenberg number, $\Wi\gg$1) regime of a startup continuous shear measurement at relatively short time/low strains, (ii) an instantaneous step strain experiment at $t=0^+$, (iii) the response to an applied stress, a creep measurement, at $t=0^+$. For the first situation, ignoring irreversible dynamics is justified for analyzing some questions, e.g, the stress overshoot, as done in the analysis of DE theory [2,3].

The study of this limit allows one to precisely define an important concept: ``microscopic absolute yielding.ÕÕ This is a criterion for an instability associated with the complete destruction of transverse localization, i.e., a mechanically-driven entanglement-disentanglement transition (EDT) [17]. In the laboratory, for displacement-controlled experiments performed at finite rates, nonlinear viscous and elastic processes occur simultaneously and compete to determine the stress-strain response. Thus, achieving absolute microscopic yield is a subtle issue that is not a priori guaranteed. For instance, relaxation decreases stresses and thus the forces on polymers that tend to destroy tube confinement. But, as we shall show in paper II, the physics associated with absolute microscopic yield underlies the predicted behavior even when the EDT is avoided.

\subsection{General Effect of External Stress}
The continuum mechanics perspective is that applied stresses generate local forces that can give rise to an affine displacement field. Strong applied forces can induce nonaffine, and ultimately irreversible, motion, which we model as a \emph{direct microscopic force} at the level of the tube field. This heuristic idea renders the problem tractable, and is not of a ``first-principlesÕÕ nature in contrast to our approach to equilibrium dynamics. It is in the spirit of the Eyring ``tilted potential energy landscape ideaÕÕ [61], but here formulated at the single polymer \emph{microscopic} force level [31,56]. This is reminiscent of a microrheology perspective, which has previously been successfully employed to treat the nonlinear viscoelasticity of colloidal glasses, colloidal gels, and polymer glasses [53,56].
   We model a constant applied stress, $\sigma$, as inducing a constant force, $f$, on the CM of every needle transverse to its axis. This results in an additional mechanical work term contributing to the tube confinement field [31-33]:
\begin{eqnarray}
F_{dyn}(r_\bot,\sigma) & = & F_{dyn}(r_\bot,\sigma=0) Ðf\cdot r_\bot \\
&=& F_{dyn}(r_\bot; \rho,S) Ð A\sigma r_\bot \approx F_{dyn}(r_\bot,\rho\sqrt{1-S})-A\sigma r_\bot \nonumber
\end{eqnarray}
Two distinct effects modify the dynamic free energy in Eq. (13). (i) Any externally induced orientational order of the rods, and (ii) a direct external force proportional to the macroscopic stress which links single rod physics (governed by the NLE evolution equation) with macroscopic rheological response. The coupling parameter is proportional to a cross sectional area, $A$, that relates the macroscopic stress and local force. In entangled rods, the plateau shear modulus is not set by the tube diameter, but rather rod orientation and length, $L$. This suggests $A\propto {{L}^{2}}$.  Two choices for the prefactor seem natural [31-33]:
\begin{eqnarray}
A & = & \pi L^2/4 \\
A &=& L^2 \nonumber
\end{eqnarray}
Predictions will be explored based on both choices. While the difference in prefactor is small, we note that quantitative features of the tube may, for some questions and some situations, control the predicted qualitative macroscopic behavior of the system. Our baseline calculations will use:
\begin{equation}
f=L^2\sigma
\end{equation}
The modest numerical prefactor uncertainty in Eq. (15) seems of little relevance compared to other simplifications implicit in Eq. (13) such as the fact that the laboratory coordinate frame, where the symmetry of the external deformation is defined, differs from the parallel/transverse coordinate system of any given rod.

Figure 3 presents results for how the characteristic length and energy scales of the anharmonic tube field change with applied stress and rod orientation. Tube dilation increases and the maximum entanglement force decreases monotonically with stress and/or orientation parameter, $S$. Note the nearly linear reduction of the maximum force with stress, and highly nonlinear tube dilation for large values of stress or $S$.
The precise definition of microscopic absolute yielding is when the external transverse force on a rod exceeds the maximum restoring force of the tube, $\ge {{f}_{\max }}$, after which the tube confinement field is completely destroyed and the dynamic free energy has no barrier to transverse displacements. This is a discontinuous transition in the sense that the transverse localization length jumps from a finite to infinite (delocalized) value as the external force crosses the critical value. The mechanical instability (force imbalance) criterion is:
\begin{equation*}
\frac{\partial {{F}_{dyn}}({{r}_{\bot }},{{\sigma }_{y}})}{\partial {{r}_{\bot }}}{{|}_{{{r}_{\bot }}={{r}_{m}}}}=0=\frac{\partial {{F}_{dyn}}({{r}_{\bot }},\rho ,S)}{\partial {{r}_{\bot }}}{{|}_{{{r}_{\bot }}={{r}_{m}}}}-A{{\sigma }_{y}}
\end{equation*}
\begin{equation}
\Rightarrow {{\sigma }_{y}}=\frac{1}{A}\frac{\partial {{F}_{dyn}}({{r}_{\bot }},\rho ,S)}{\partial {{r}_{\bot }}}{{|}_{{{r}_{\bot }}={{r}_{m}}}}=\frac{{{f}_{\max }}(\rho ,S)}{A}
\end{equation}
This criterion can be expressed in terms of a critical stress or force on the rod CM, with inter-conversion via Eq. (15). 

\begin{figure}
\centerline{\includegraphics[width=0.9\linewidth]{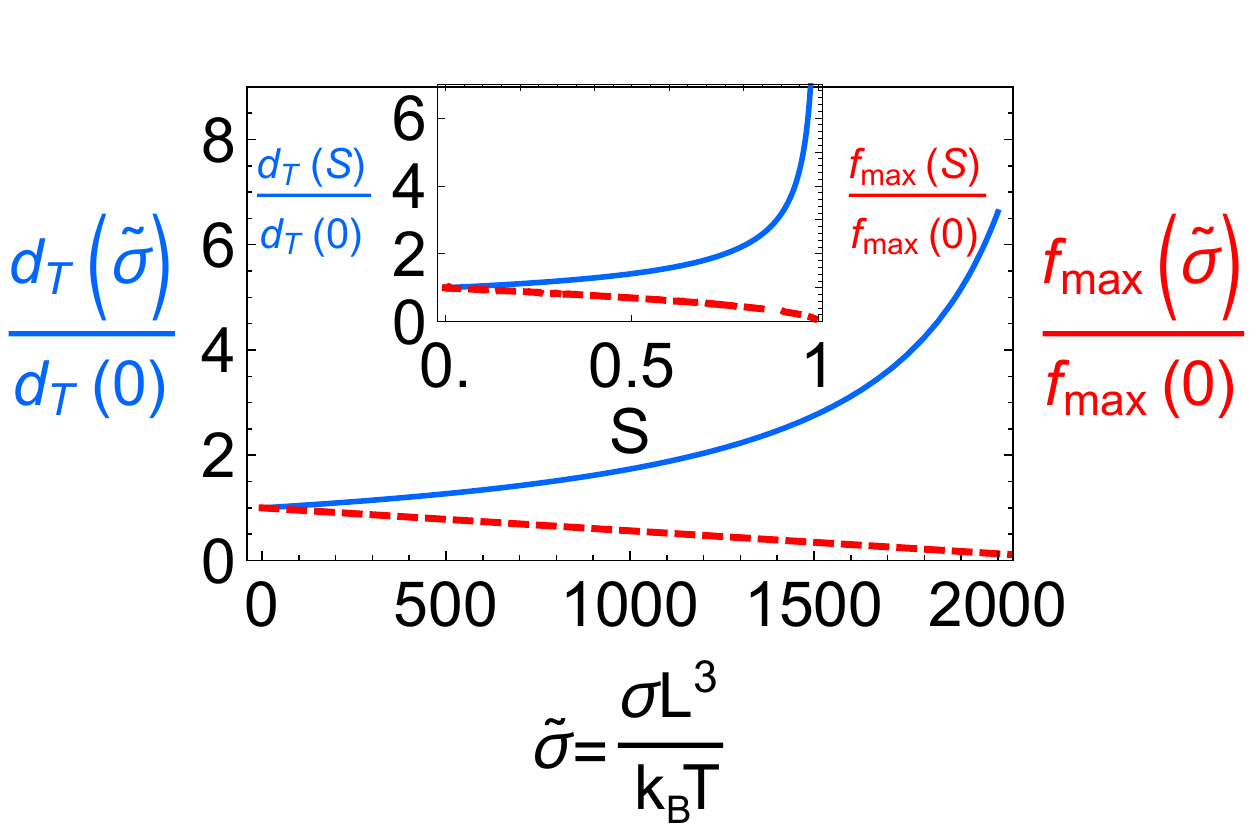}}
\caption{Variation of key length and force scales of the tube confinement potential with applied dimensionless stress and degree of needle orientation.  Main frame shows how the mean tube diameter and maximum confining force vary with stress. Inset shows how the same two quantities vary with the nematic orientational order parameter. For both the main frame and the inset, the solid blue curves correspond to the left vertical axis (normalized tube diameter), and the red dashed curves correspond to the right vertical axis (normalized maximum entanglement force).}
\end{figure}

\subsection{Implications of Finite Tube Confinement Strength: Microscopic Yielding}
Recall for rods that the elastic shear modulus, $G_e$, is not dependent on the tube diameter, but rather reflects slow orientational stress relaxation of a rod in a tube [2,57]: 
\begin{equation}
{{G}_{e}}=\frac{3}{5}\rho {{k}_{B}}T
\end{equation} 	
If the tube is broken this entropic elasticity is lost, reptation becomes irrelevant, and orientational stress rapidly decays. Thus, a rheological stress overshoot \emph{could} be a signature of tube breaking or massive softening. In the regime of present interest, a force on the rod can be estimated from a nonlinear elastic equation-of-state associated with macroscopic affine deformation, $\sigma(\gamma)$. This discussion frames the microscopic absolute yield problem as a point of \emph{imbalance} between the external and the intrinsic tube forces.

\subsubsection{Stress-Controlled Force Imbalance}
The microscopic tube confinement force is given in Eq. (11). Using Eq. (15), and equating it to an applied stress ($\sigma$), defines the microscopic absolute yield stress, $\sigma_y$,
\begin{equation}
{{\sigma }_{\max }}=\frac{{{f}_{\max }}}{A}\approx \frac{{{f}_{\max }}}{{{L}^{2}}}=4\frac{{{k}_{B}}T}{{{d}_{T}}{{L}^{2}}}\equiv {{\sigma }_{y}}
\end{equation} 

Using Eqs. (4) and (17), this can be expressed in units of the linear modulus as:
\begin{equation}
\frac{{{\sigma }_{y}}}{{{G}_{e}}}=\frac{5\pi }{12\sqrt{2}}\approx 0.93
\end{equation}
This ratio can be interpreted as an apparent ``microscopic yield strainÕÕ, ${{\gamma }_{y,abs}}$, which (in the heavily entangled limit) is independent of rod density. An applied stress of this magnitude or higher destroys tube localization, implying a fragile entanglement network that can be destroyed by strains of order unity or stresses of order the equilibrium shear modulus.

If $A=\pi {{L}^{2}}/4$, the apparent yield strain is modestly larger, $\gamma_{y,abs}\sim 1.18$. A similar modest numerical ambiguity concerns the fraction of force from the macroscopic stress that acts transverse to the rod axis. If, e.g., the fraction is one half, then we predict $\gamma_{y,abs}\sim 2 Ð 2.4$. All our yield strain estimates are roughly equal to, or roughly a factor of two less than, the strain value of the stress overshoot predicted by DE theory [2] ($\sim 2.25$) based on the classic assumption of an infinitely strong, affine-deforming tube. This raises a subtle but fundamental question: is the stress overshoot in startup shear due to the DE mechanism, the tube destruction/weakening mechanism, or a combination of both? Determining the answer seems subtle given the above estimates of the absolute yield strain and the relevance of numerical prefactors of order unity in our non-classical microscopic yield mechanism.
 	For heavily entangled systems, we derived the following approximate expression which captures the stress- and orientation-driven pre-yielding tube dilation and is valid for $S$ not too close to unity and for stresses not too close to the yield stress [32,33]:
\begin{equation}
\frac{{{d}_{T}}(S,\sigma )}{{{d}_{T}}}\approx \frac{1}{\sqrt{1-S}-\frac{\sigma {{L}^{3}}}{3{{k}_{B}}T}\frac{{{\rho }_{e}}}{\rho }}\ =\frac{1}{\sqrt{1-S}-\frac{{{\rho }_{e}}{{L}^{3}}}{5}\cdot \frac{\sigma }{{{G}_{e}}}}
\end{equation}

\subsubsection{Strain-Controlled Force Imbalance}
Given a relation between stress and strain, the analysis in the preceding section is immediately relevant to the question of microscopic absolute yielding after a step strain, or as a limiting estimate of the stress overshoot in a startup shear for fast flows. The simplest calculation adopts a linear stress-strain curve: 
\begin{equation}
\sigma = G_e\gamma
\end{equation}
Using this in Eq. (15) leads to the identical apparent yield strain values discussed above.
  Going beyond the na•ve calculation requires including the classic orientation effects on the elastic shear stress projection in the stress-strain expression [2,3]. Note that for rods the tube diameter does not enter the plateau modulus, nor does the relevant cross-sectional area, $A$. Thus, one has:
\begin{equation}
\sigma ={{G}_{e}}\gamma h(\gamma )\approx {{G}_{e}}{{\frac{\gamma }{1+{{\gamma }^{2}}/5}}_{{}}}
\end{equation}

where the second relation defines the ``damping functionÕÕ [2,3], $h(\gamma)$. But the intrinsic tube cohesion weakens due to the predicted tube dilation, per Eq.(20). This effect is not based on affine tube deformation; the affine deformation induces rod \emph{orientational order} that in turn dilates the tube [29]. The resulting tube diameter is [29]:
\begin{equation}
\frac{{{d}_{T}}(\gamma )}{{{d}_{T}}}\cong \frac{1}{\sqrt{1+{{\gamma }^{2}}/4}}\approx \frac{1}{\sqrt{1-S}}
\end{equation}
The second approximate equality follows from the classic Lodge-Meissner relation [2], 
\begin{equation}
S(\gamma )=2\gamma {{\left( 3\sqrt{4+{{\gamma }^{2}}}-\gamma  \right)}^{-1}}\ \ \textrm{for}\ \ S>0
\end{equation}
Note that Eq. (23) includes only affine deformation physics, not the direct force effect present in Eq. (20) above.
We now perform two calculations to establish the influence of tube dilation and stress over-orientation on microscopic absolute yielding. First, tube dilation is included (lowering yield strain), and a linear stress-strain relation is employed. By combining Eqs.(18) and (23) one obtains: 
\begin{equation}
{{\sigma }_{y}}=\frac{5\pi }{12\sqrt{2}}{{G}_{e}}\cdot \frac{1}{\sqrt{1+\gamma _{y,abs}^{2}/4}}={{G}_{e}}{{\gamma }_{y,abs}}
\end{equation}
Solving this yields:
\begin{equation}
\gamma_{y,abs} \approx 0.85
\end{equation}
This is a small $\sim 10\%$ reduction relative to the result obtained using Eq. (15). If one employs $A=\pi {{L}^{2}}/4$, then $\gamma_{y,abs}\sim 2 Ð 1.04$.  If the $h$-function in Eq. (22) is included in addition to the tube dilation effect, this reduces the tube softening force. Solving Eq. (25) with the h-function on the right hand side does lead to the prediction of microscopic yielding, albeit with a modestly larger yield strain and stress: 
\begin{equation}
\gamma_{y,abs}\approx 1
\end{equation}
This is almost identical to the na•ve calculation result in Eq. (19) due to a near compensation of the tube dilation and damping function effects. Overall, the conclusion that the yield strain is of order unity appears to be robust.
The microscopic absolute yield strains and stresses computed above reinforce our central message that the entanglement network is mechanically fragile and can be destroyed in the nonlinear elastic scenario. In a displacement-controlled step-strain or startup shear experiment, if the tube breaks the stress will rapidly drop, which would in turn leads to ``healingÕÕ or re-emergence of the entanglement network. A full numerical nonlinear step strain rheological calculation has been performed [32], which showed tube breaking followed by healing with distinctive kinetics. In an ideal creep experiment, an applied stress beyond the microscopic absolute yield value would completely destroy the transverse dynamical constraints. In the real world, such tube breaking events may lead to macroscopic instabilities such as melt fracture, wall slip and shear banding, which are not addressed by our approach.

\section{Relaxed Primitive Path Description of Entangled Chain Polymer Melts}
We now discuss the consequences of extending our needle fluid results to chain liquids within a self-consistent disconnected primitive path (d-PP) framework. This analysis is more sensitive to numerical prefactors associated with transmission of macroscopic stresses to microscopic forces and other approximations of the theory.

\subsection{Mapping from Rods to PP Chains in Equilibrium}
The classic DE picture of a flexible polymer equilibrated in a tube is a freely jointed chain of rigid PP segments of step length, $L_e$. A key assumption is contour length equilibration. Under equilibrium and deformed conditions, this picture implies [3]:
\begin{equation}
{{L}_{e}}=\kappa 2{{r}_{loc}}\equiv \kappa {{d}_{T}}=\kappa \sigma \sqrt{{{N}_{e}}}
\end{equation}
In the DE approach [2], there is only one entanglement-related length scale corresponding to $\kappa =1$ in Eq. (28). However, for real chain liquids, the most elementary description involves transverse displacements of monomers or beads/segments relative to the coarse-grained PP step, and simulations of entangled polyethylene and polybutadiene melts find $\kappa \approx 2.5$ [36]. Our present theory takes the relationship between $L_e$ and $d_T$ as input, and below we make predictions based on $\kappa =1$ and $\kappa =2.5$. This is one of several quantitative ÒprefactorÓ issues that enter our analysis of chain liquids based on a coarse grained PP description. 
Adopting Eq. (28), our self-consistent relation for the (Gaussian, harmonic-fluctuation level) tube diameter or PP step length is derived to be [28,29]:
\begin{eqnarray}
1 & = & \frac{\rho_{PP}L_e^3}{16 \pi^2\sqrt{2}} F\left ( \frac{2 L_e}{d_T} \right)G \nonumber \\
F(x)&=& \frac{8\sqrt{2}}{\pi \rho L}\beta K_\bot (x) \\
G &\equiv & \int d\vec{u}_1 d\vec{u}_2 \sqrt{1- (\vec{u}_1\cdot \vec{u}_2 )^2} \alpha (\vec{u}_1) \alpha (\vec{u}_1) \nonumber 
\end{eqnarray}
As discussed in Section II, $F(x)$ carries information about the strength of the confining entanglement forces (see Eqs.(8) and (9)), and $G$ quantifies the effect of any PP orientational order on the tube diameter [29]. Here, as before, affine tube deformation is not assumed. Instead, the orientational order induced from the global strain is assumed to be transmitted to the PP level; the theory then predicts the tube diameters changes in a non-affine manner, the details of which are different under shear and extension. If Eq. (28) is then adopted (an assumption that will be relaxed in section V), Eq. (29) is a closed equation for the tube diameter or PP step length. 

To compute orientational stress, the IAA idea of DE is adapted to our framework; specifically, we model the PP chain liquid as a fluid of disconnected PP (d-PP) needles for the purpose of tube scale physics and orientational stress response. For questions involving reptation, global connectivity is retained, e.g., when computing the reptation or disentanglement relaxation time [28]. Based on this mapping, the entanglement shear modulus relevant to orientational stress stored by chains in their tubes follows from DE [2] as:
\begin{eqnarray}
{{G}_{e,chain}} &\to& \frac{3}{5}{{\rho }_{PP}}{{k}_{B}}T={{\rho }_{chain}}\frac{N}{{{N}_{e}}}\frac{3}{5}{{k}_{B}}T=\frac{3}{5}\frac{{{\rho }_{seg}}{{k}_{B}}T}{{{N}_{e}}} \nonumber \\
& =& \frac{3}{5}{{\rho }_{seg}}{{k}_{B}}T{{\left( \frac{\sigma }{{{d}_{T}}} \right)}^{2}}
\end{eqnarray} 
Equation (30) is corrected for contour length fluctuations by simply changing [2] the prefactor $3/5 \rightarrow 4/5$. More generally, the precise prefactor touches on the question of the partition of entanglement stress storage into stretch and orientation components, a subtle quantitative issue. We model this by introducing a ``stress partitioningÕÕ parameter, $\lambda$:
\begin{equation}
{{G}_{e,chain}}=\lambda \frac{{{\rho }_{seg}}{{k}_{B}}T}{{{N}_{e}}}
\end{equation}
plausible extremes of which are $1/2$ and 1.  Full resolution of the proper value of $\lambda$ awaits a deeper theory that treats chain stretch and retraction in a first-principles manner.

Recall that the PP step length in Eq. (29) is not a priori assumed, it is self-consistently derived. For equilibrium fluids, based on $\kappa = 1$ in Eq.(28), the PP level tube diameter is predicted to be:
\begin{equation}
d_{T,PP}\approx 10.2p
\end{equation}
where $p$ is the packing length [62,63]. If one employs $\kappa =2.5$, we find ${{d}_{T,PP}}\approx 8.3p$. Our results for the tube diameter are relatively insensitive to $\kappa$ as long as it is greater than or equal to unity [29], and we believe values of $\kappa <1$ are unphysical.
PP analysis of simulations provide the ``exactÕÕ result to compare with our theoretical results [64]: 
\begin{equation}
d_{T,PP}\approx 12.2p
\end{equation}
Rheological experiments that empirically extract a tube diameter using a model find [6]:
\begin{equation}\tag{33b}
d_{T,PP}\approx 17.7p
\end{equation}
and the correlation with the plateau shear modulus [6,62]:
\begin{equation}
{{G}_{e}}\approx 0.0023\frac{{{k}_{B}}T}{{{p}^{3}}}\approx 12.75\frac{{{k}_{B}}T}{d_{T}^{3}}
\end{equation}  

We have analyzed the full PP step length distribution function from the dynamic free energy predicted by the d-PP mapped theory [28]. It quantitatively agrees with the distribution extracted from simulations of entangled chain melts for $\kappa  =2.5$ [36]. Choosing $\kappa = 1$ leads to qualitative agreement. Significantly, we find quantitative agreement using the $\kappa$value that is consistent with that deduced from simulation. On the other hand, our prediction for the tube diameter, which is a harmonic property of the dynamic free energy, is a bit closer to that deduced from PP simulation analysis if one chooses $\kappa = 1$. Curiously, as our theory predicts, the basic functional form of the PP distribution function for simulated chain melts and the transverse displacement probability distribution for experimental F-actin systems do seem very similar [28].  
	All of the effects of affine orientation and strain on tube diameter derived for rods carry over to d-PP chains. The finite tube strength is [28,29]: 
\begin{eqnarray}
f_{max} & \approx & \frac{k_B T}{d_{T,PP}},\quad \kappa = 1 \nonumber \\
& \approx & 1.6  \frac{k_B T}{d_{T,PP}},\quad \kappa = 2.5
\end{eqnarray}
This maximum force is the crucial anharmonic property of the tube field, and is quantitatively sensitive to $\kappa$. Moreover, it is weaker, by a factor of $\sim 2.5 Ð 4$  than predicted for heavily entangled rods in Eq. (11). Given Eqs. (32) and (33), the connection between our predicted PP tube diameter and the extracted rheological one is:
\begin{eqnarray}
d_{T,rheo} \equiv d_T & \approx &1.73 d_{T,PP}, \quad \kappa = 1 \nonumber \\
& \approx & 2.13 d_{T,PP}, \quad \kappa = 2.5
\end{eqnarray}
The effect of deformation on the tube diameter is comparable in magnitude to an affine deformation as found for rods, and the tube dilation Eq. (23) applies.

\subsection{Microscopic Absolute Yielding}
We now carry out the Ònonlinear elasticÓ analysis of force imbalance for PP chain melts, as we did for rods in section III. Its relevance to startup shear experiments is at best for the heavily entangled, slow flow, DE-like regime where Wi>>>1 but the Rouse Weissenberg $\textrm{Wi}_R \ll 1$ corresponding to contour length equilibration on strain scales small compared to, e.g., the overshoot region of the response. Doi-Edwards theory predicts under these conditions that the overshoot occurs at ${{\gamma }_{m}}\approx \sqrt{5}$, the same strain value as for rods since the mechanism is the same affine over-orientation effect. The question is whether our theory predicts a different physical scenario for a PP-level description of chains, as it does within the microscopic yielding scenario for rods. The analysis differs in detail from that for rods for several reasons, including the fact that the entanglement modulus now involves the tube diameter. This implies the physically relevant cross sectional area to convert force to stress is the entanglement mesh size (tube diameter). A simple assumption is:
\begin{equation}
A\approx \frac{\pi}{4}d_T^2
\end{equation}

We first consider the stress-controlled situation. From the d-PP mapping, one finds:
\begin{equation}
{{\sigma }_{\max }}=\frac{{{f}_{\max }}}{A}\approx \frac{{{f}_{\max }}}{\pi d_{T,PP}^{2}/4}=\frac{4}{\pi }\left( 1+\frac{3}{5}{{\delta }_{\kappa ,2.5}} \right)\frac{{{k}_{B}}T}{d_{T,PP}^{3}}\equiv {{\sigma }_{y}}
\end{equation}
Using our results for the PP tube diameter gives:
\begin{eqnarray}
\sigma_y & = & \frac{1}{12.75}\frac{4}{\pi} (1+\frac{3}{5}\delta_{\kappa,2.5} )G_e \frac{d_T^3}{d_{T,PP}^3} \nonumber \\
& = & \frac{1}{12.75}\frac{4}{\pi} (1+\frac{3}{5}\delta_{\kappa,2.5} ) (1.73+\frac{2}{5}\delta_{\kappa,2.5} )^3 G_e \equiv \gamma_{y,abs}G_e \nonumber \\
\gamma_{y,abs}&\approx & 0.5\ (1.5),\quad \kappa = 1\ (2.5)
\end{eqnarray}
Here we have used the Kronecker delta notation only to distinguish our predictions for $\kappa =1.0$ and $\kappa =2.5$. Given that $\kappa =2.5$ leads to quantitative agreement of our theory with simulations for the non-Gaussian aspects of the full PP distribution function [28, 36], it is the superior choice for analyzing microscopic yielding. The apparent absolute yield strain defined above is not dependent on degree of entanglement. Its absolute magnitude is below the DE value of $\sqrt{5}$, though only modestly so for the $\kappa =2.5$. 

We next repeat the displacement- (strain) controlled analysis of Section III to establish the influence of the damping h-function and tube dilation on the microscopic yield criterion. \emph{To do this requires knowing the relation between the tube diameter,} $N_e$, \emph{and} $G_e$ \emph{under deformation, a non-trivial, open problem in polymer physics.} If we literally follow the rod analysis, then orientation dilates the tube, thereby weakening the maximum confining force, and the damping function weakens the direct force, but there is no change of $G_e$ due to deformation. However, based on the poor understanding of what entanglements are and how they change under deformation, other approaches are possible. For example, one could adopt Eq. (22) literally, which ignores any strain-induced tube dilation or softening of the entanglement restoring force in the dynamic free energy. To compute the external force, one would thus assume no change of $G_e$ with strain, corresponding to the physical picture that $N_e$ (and $Z=N/N_e$) is strain-invariant as in a simple treatment of crosslinked rubber. One then obtains:
\begin{equation}
{{\sigma }_{y}}\approx {{\gamma }_{y,abs}}{{G}_{e}}=\lambda {{G}_{e}}\frac{{{\gamma }_{y}}}{1+\gamma _{y}^{2}/5}
\end{equation} 
where ${{\gamma }_{y,abs}}\approx 0.5\ (1.5)$ for $\kappa =1(2.5)$, per Eq. (39). Solving the quadratic equation, we find that for $\ \kappa =2.5$ there is no physical solution for any value of $\lambda$, and hence absolute microscopic yielding is not predicted. On the other hand, for $\ \kappa =1$, corresponding to a weaker maximum entanglement force, tube breakage is predicted at a strain that varies from $\sim 0.5 - 1.4$ for $\ \lambda =1-0.5$. Thus, in this first approach the qualitative issue of microscopic yield depends on the quantitative strength of the tube.

The above analysis assumed a confinement field invariant to strain, which is not plausible in the context of our theory. We now consider a second approach which we believe is far more realistic since it takes into account our predicted strain softening of the tube diameter and maximum confinement force. The mechanical modulus Ge remains unchanged, corresponding to the rubber-like assumption of conserved number of PP steps. To be consistent, we expect the cross-sectional area required to convert a macroscopic stress to a microscopic force should change if the tube diameter changes. Adopting this assumption, one obtains: 
\begin{equation}
{{\sigma }_{y}}\approx \frac{{{\gamma }_{y,abs}}}{{{\left( 1+\gamma _{y}^{2}/4 \right)}^{3/2}}}{{G}_{e}}=\lambda {{G}_{e}}\frac{{{\gamma }_{y}}}{1+\gamma _{y}^{2}/5}
\end{equation}
This self-consistent equation has a solution for all sensible parameter values. For
$\kappa = 1$ we find:
\begin{equation}
\gamma_y \approx 0.9 Ð 0.5 \ \textrm{for}\ \lambda = 0.5 - 1
\end{equation}
and for $\kappa=2.5$ we find 
\begin{equation}\tag{42b}
\gamma_y \approx 2.0 Ð 1.2 \ \textrm{for}\ \lambda = 0.5 - 1
\end{equation}
Thus, self-consistently including the tube softening effect results in microscopic absolute yielding. The ``yield strainsÕÕ vary significantly from $\sim 0.5 Ð 2$, but based on the more realistic value of  $\kappa = 2.5$ they are comparable to low-$\Wi_R$ measurements of the stress overshoot for entangled chain melts (strains $\sim 1.6 Ð 2$) [65]. The numbers in Eq. (42) are also comparable to the predicted ÒapparentÓ absolute microscopic yield strains in Eq. (39) which were deduced in the stress scenario. Overall, these results suggest that it is plausible that the observed stress overshoot in chain melts at low $\textrm{Wi}_R$ \emph{might} be fundamentally related to tube destruction, or at least massive softening, and not simply to affine over-orientation. This scenario is seems akin, at least in spirit, to the ``elastic yieldingÕÕ concept of Wang et al. [17-19,22].  One can imagine other assumptions can be invoked in the microscopic yield calculation. But we believe it is physically incorrect to assume $G_e$ changes with strain, or that $A$ is strain invariant. Thus, our intuition is that the second approach above is the more physical one. 

The above results may be of relevance for entangled chain melts at low $\textrm{Wi}_R$ , and several comments are in order. (i) From a theoretical perspective, the idea that the stress overshoot is due to, or is significantly influenced by, massive softening of the entanglement field is qualitatively new. However, a definitive conclusion is difficult since our non-classical mechanism is sensitive to numerical prefactors of order unity. (ii) Per entangled rods, we can treat microscopic yielding under ideal step-strain and creep conditions, compute the tube survival function, etc., for entangled chain melts at the d-PP chain level. (iii) We find that the quantitative value of the numerical prefactors that enter our theory can lead to qualitatively different behavior. This seems unavoidable since the precise value of the maximum tube confining force is critical in determining whether a force imbalance can occur. Such a situation is reminiscent of the quantitative sensitivity of the GLaMM model [7,14], where if the CCR strength is too weak, an unphysical flow curve is predicted, but if it is too strong the stress overshoot prediction can be degraded.

\section{Tube Diameter and Microscopic Yielding for Stretched Chains}
	We now consider the opposite limit where stretched chains do not relax at all on the time scale of interest. In this scenario, deformation induces changes of both orientation and contour length. We explore how this modifies the tube diameter and the ability of an external force to destroy the tube in the nonlinear elastic scenario. 

\subsection{General Formulation and Tube Diameter in Shear}
To proceed, three major issues must be addressed. (i) How are the 3 length scales in Eq. (28) related under deformation? (ii) How, if at all, does the plateau modulus, $G_e$, change? (iii) Can a full 3-d tensor treatment of the confinement field be performed given there is a ``tube length scaleÕÕ in multiple directions (dictated by the symmetry of the external deformation)?  These are open and difficult questions. For (iii), we proceed as before and pre-average the tensorial aspects to compute a single mean tube diameter. The GLaMM model [7] acknowledges the above complexities, and the uncertainty concerning how to address them. In practice, GLaMM assumes [7]: (a) the tube diameter does not change with deformation, (b) the number of entanglements $Z=N/N_e$ grows with deformation in proportion with the chain PP length and hence the effective $N_e$ decreases, (c) $G_e$ does not change.  Different alternative plausible scenarios were noted, including the idea that the number of entanglements is unaffected by deformation and that the tube diameter changes in a nonaffine manner [7].

Here, for the time scales of interest in this paper, we invert assumptions (a) and (b), and adopt (c). Thus, the number of PP segments, or of entanglements $Z$, does not change upon stretching, and $N_e$ is fixed at its quiescent value. The latter assumption is a zeroth order model for a crosslinked rubber, and implies $G_e$ does not change; we view it as the simplest scenario if one subscribes to the transient network model picture of an entanglement network, and a natural assumption on time scales short relative to stretch relaxation. 

The chain of PP strands is then modeled as a biased random walk with strain-dependent ``longitudinal'' and ``transverse'' step lengths of: 
\begin{align}
  & {{L}_{e}}\equiv {{f}_{||}}(\gamma )\cdot \kappa \sigma \sqrt{{{N}_{e}}} \nonumber \\ 
 & {{d}_{T}}\equiv {{f}_{\bot }}(\gamma )\cdot \sigma \sqrt{{{N}_{e}}} 
\end{align}
We analyze the harmonic tube diameter question based on the simplest model where $\ \kappa =1$; recall that the predicted tube diameter changes only by $\sim 20\%$ as $\ \kappa =1 \rightarrow 2.5$ [28]. The self-consistent transverse localization of Eq.(29) can then be written as 
\begin{equation}
\frac{16{{\pi }^{2}}\sqrt{2}}{F\left( \frac{2{{f}_{||}}(\gamma )}{{{f}_{\bot }}(\gamma )} \right)\ G(\gamma )}=\frac{{{d}_{T,0}}}{p}f_{||}^{3}(\gamma ) 		
\end{equation}
where $d_{T,0}$ is the equilibrium tube diameter and $G(\gamma )$ the orientation factor that results from an affine deformation tensor at the global level being transmitted to the PP scale [29]. A natural choice for the parallel direction is an affine stretch, which for shear is:
\begin{equation}
{{f}_{||}}(\gamma )=\sqrt{1+{{\gamma }^{2}}/3}
\end{equation}
For the transverse direction, we view the most natural guesses as either a Òvolume conservingÓ ansatz or no change:
\begin{equation}
{{f}_{\bot }}(\gamma )=\frac{1}{\sqrt{{{f}_{||}}(\gamma )}}\quad \textrm{or} \quad 1
\end{equation}
On the scales of the plots presented, our numerical predictions are insensitive to which choice is adopted, as the effect of the ${{f}_{||}}(\gamma )$ dominates the changes in the tube diameter.

Using Eqs. (45) and (46) in Eq. (44) yields a closed equation. The numerical results for a step shear strain are shown in Figure 4, and are compared with our prior results based on rapid stretch relaxation. The dramatic conclusion is that chain stretch without relaxation induces a strong \emph{compression} of the tube diameter. Qualitatively, the tendency of chain stretch (by itself) to enhance tube localization is evident from the PP number density factor ${{\rho }_{PP}}L_{e}^{3}$ in Eq. (29). We find that our numerical calculations are well approximated by the analytic form: 
\begin{equation}
\frac{{{d}_{T}}(\gamma )}{{{d}_{T}}}\approx \frac{\sqrt{1+{{\gamma }^{2}}/4}}{1+{{\gamma }^{2}}/2}
\end{equation}
The numerator reflects orientation-driven dilation and the denominator reflects tube compression due to stretching. For asymptotically large strains one has:
\begin{equation}
\frac{{{d}_{T}}(\gamma )}{{{d}_{T}}}\approx {{\gamma }^{-1}}\quad ,\quad \gamma \gg1
\end{equation}
Net tube compression implies an effective ``tighteningÕÕ of entanglements in the \emph{dynamical} (not elastic modulus) sense that reptation-driven disentanglement slows down. Curiously, although the GLaMM model [7] assumes the tube diameter is unchanged under deformation, its assumption that the number of entanglements grows also leads to the idea that the reptation rate is suppressed when chains become stretched.  

 	The above results suggest a novel scenario whereby the tube diameter and disentanglement rate are non-monotonic functions of time (or accumulated strain) in a startup continuous shear deformation. At early times when chains are stretched the tube diameter is compressed and disentanglement is slowed down relative to the equilibrium behavior. However, when chains retract, there will be a crossover back to tube dilation and accelerated reptation and disentanglement. Intriguing evidence for such a behavior has been found in recent Brownian dynamics simulations [24], although the crucial issue of the time scale of stretch relaxation remains to be definitively settled [14,24,25].

\begin{figure}
\centerline{\includegraphics[width=0.9\linewidth]{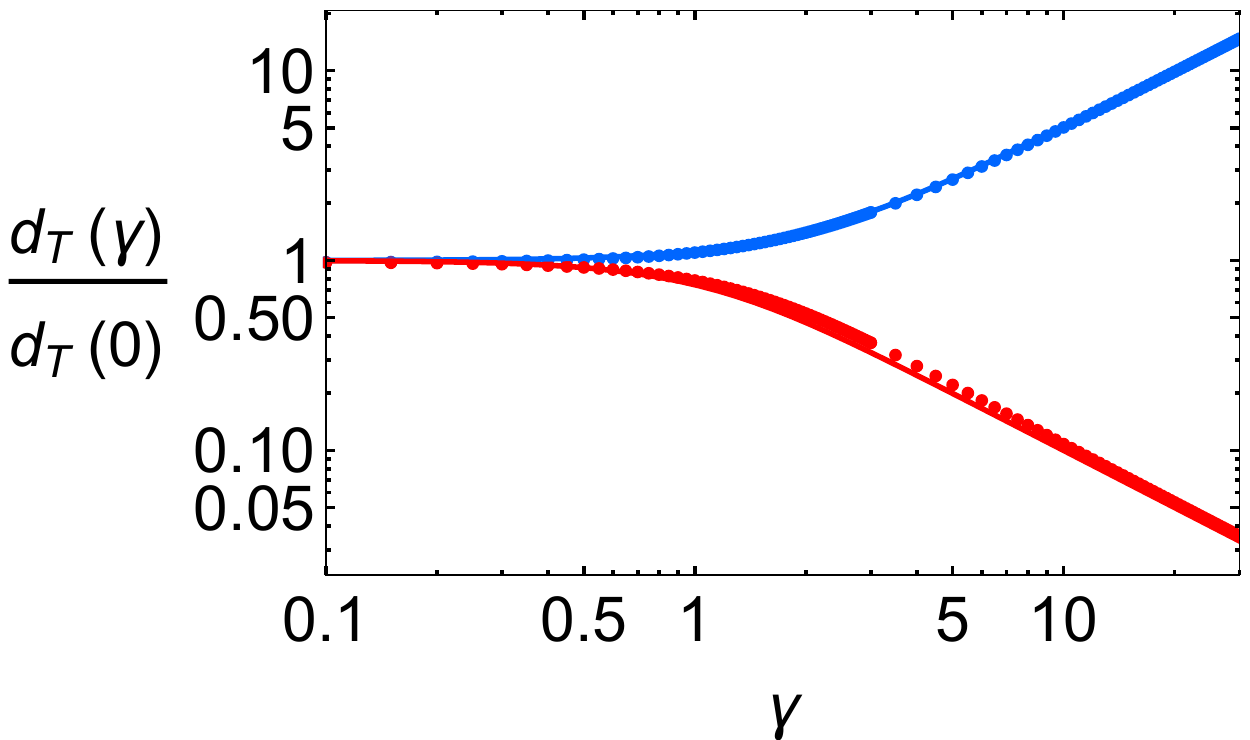}}
\caption{Log-log plot of the tube diameter of a chain fluid subjected to a globally affine shear deformation, normalized by its equilibrium value, as a function of shear strain.  The upper (lower) results correspond to unrelaxed affinely deformed (relaxed contour length) rigid needle primitive path chains. The solid curves through the numerical points are the analytic functions in the text, Eq. (47) and Eq. (21), respectively.}
\end{figure}

\subsection{Tube Diameter in Uniaxial Extensional Deformation}
We have also performed the stretched chain analysis for an extensional
deformation of amplitude $\lambda$. The longitudinal step length is of the standard affine form [2],
\begin{equation}
\frac{{{L}_{e}}(\lambda )}{{{L}_{e}}(1)}=\frac{\lambda }{2}\left[ 1+\frac{{{\sinh }^{-1}}\left( \sqrt{{{\lambda }^{3}}-1} \right)}{\sqrt{{{\lambda }^{3}}({{\lambda }^{3}}-1)}} \right]
\end{equation}
The transverse factor is given by Eq.(46), and the numerical results below are again dominated by the parallel step length function. The transmission of global orientational order induced by the affine transformation to the PP level is again assumed [29] in our computation of $G(\lambda )$. With this, Eq. (44) allows the mean tube diameter to be computed.

Results are shown in Figure 5. All qualitative behaviors are the same as predicted under shear, including strong tube compression. At large deformations, the tube diameter ratio scales as ${{\lambda }^{-1/2}}$ and ${{\lambda }^{3/2}}$ with the inclusion of chain stretch or not, respectively.  

\begin{figure}
\centerline{\includegraphics[width=0.9\linewidth]{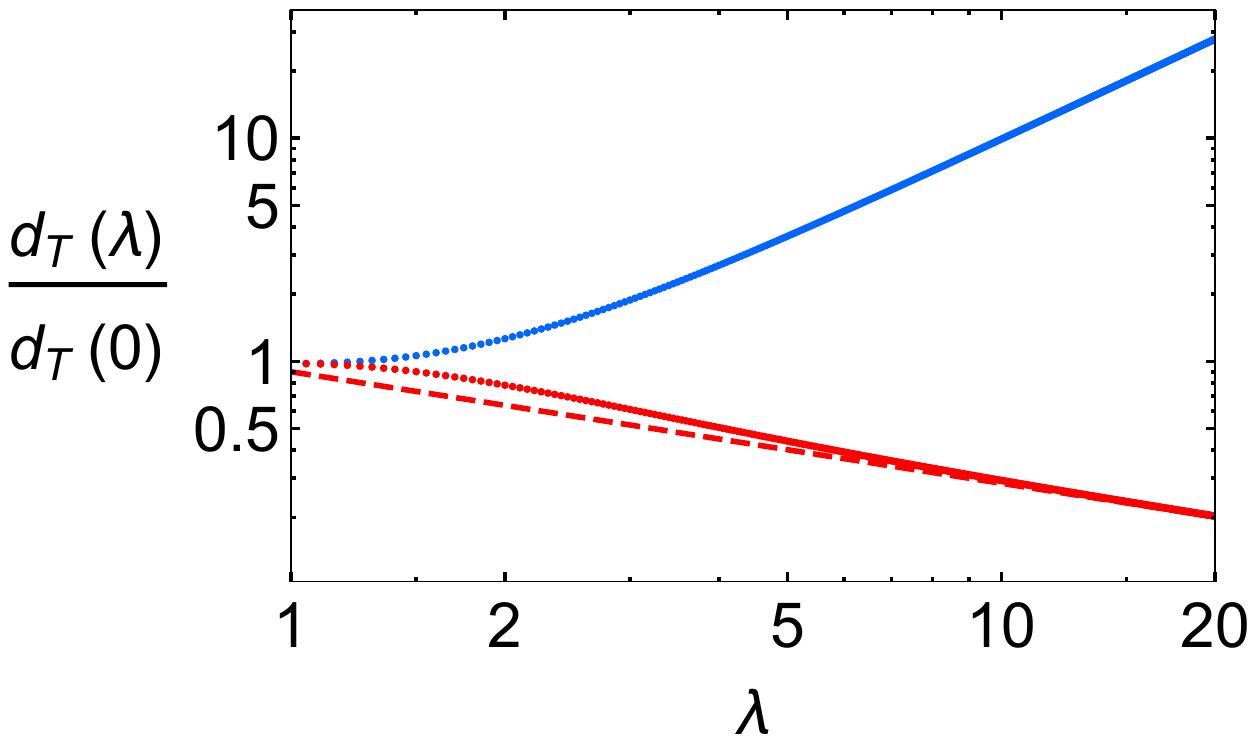}}
\caption{Tube diameter of a chain fluid subjected to a globally affine extensional deformation, normalized by its equilibrium value, as a function of extension ratio. The lower curve corresponds to the contour-length-relaxed PP-level calculation and the upper curve assumes no relaxation of chain stretch. The red dashed curve is the large deformation analytic scaling result discussed in the text.}
\end{figure}

Figure 6 compares shear versus extension tube diameter predictions based on common measures of extensional deformation: Henky strain, $\log (\lambda)$, or a rubber-like form, $\lambda - \lambda^{-2}$. Tube compression, and thus entanglement tightening, is stronger in extension compared to shear based on Hencky strain, but there is no clear trend based on the rubber-like strain measure. Stronger tube compression corresponds to greater suppression of dynamical disentanglement, and thus a greater possibility of rubber-like response if chains do not retract. Such an inference might be consistent with the non-classical argument of Wang and coworkers[66] that ``strain hardeningÕÕ is a nonlinear elastic effect.

\begin{figure}
\centerline{\includegraphics[width=0.9\linewidth]{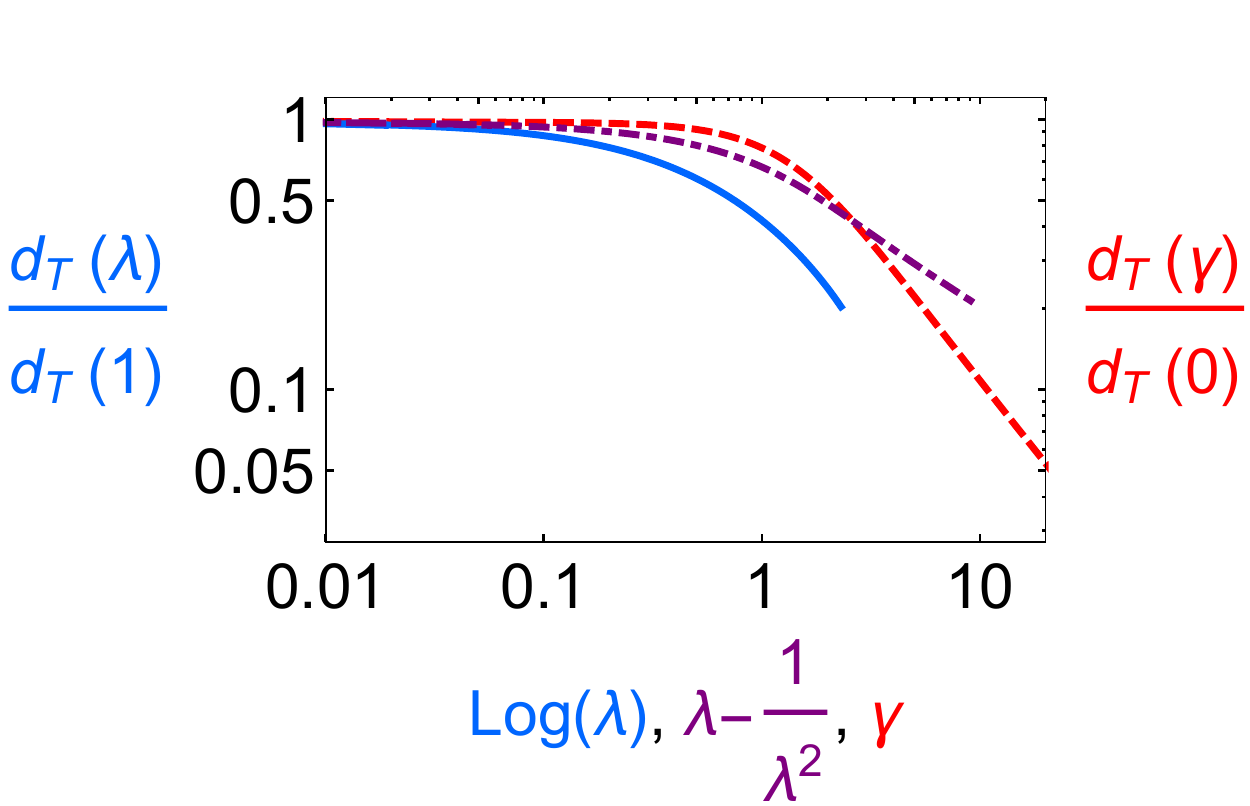}}
\caption{Comparison of the unrelaxed stretched chain predictions for how the tube diameter changes under shear (red dashed curve) and extension. Two measures of extensional strain are used (see text): a Hencky strain (solid blue curve) and an alternate rubber-like measure (dash-dotted purple curve).  }
\end{figure}

\subsection{The deGennes Non-classical Proposal}
In 1991, motivated by the arguable at the time (and still controversial [14,17,23-25]) issue of whether deformed chains do or do not freely retract in a Rouse manner under strong shear deformation, deGennes [67] speculated in a short paper that stretched chains in fact do not freely retract. As a consequence, he argued that the tube diameter is compressed as the inverse linear power of shear strain. This result is in qualitative accord with our microscopic predictions discussed above. However, de Gennes arrived at this conclusion via theoretical arguments very different than ours. Specifically, he invoked the Pincus tension blob idea within an elastic stress-strain framework, together with a non-mechanical argument for the lack of free retraction based on a chain trapped in a capillary (assuming the sub-strand-scale conformations remained an ideal random walk). By assuming chain stretch did not change the elastic modulus (as per our assumption of conserved $N_e$ or $Z$), deGennes predicted tube compression, including the scaling law in Eq.(48). However, the underlying physics of deGennesÕ and our approach seem very different. Indeed, as a consequence of the assumptions of his analysis, deGennes argued [67] the shear viscosity of entangled chain polymers does not shear thin for a monodisperse melt! This seems wrong, and qualitatively disagrees with our theoretical predictions [33].

\subsection{Microscopic Absolute Yielding}

We now determine how unrelaxed chain stretching influences microscopic absolute yielding under shear. We know from sections II and III this question is sensitive to how the tube diameter, entanglement modulus and maximum entanglement restoring force evolve with strain. It also can be sensitive to numerical prefactors. We follow our earlier analyses, and investigate what we predict if one assumes a linear stress-strain relation consistent with the crosslinked rubber analogy for unrelaxed stretched chains. 

The tube confinement field instability condition of Eq. (41) is modified for unrelaxed chains. There is no stress softening factor due to affine over-orientation (the h-function is absent) on the right hand side (RHS) of Eq. (41), and $G_e$ is invariant to stretching. Thus, the RHS is of an ideal Hookean rubber form. The left hand side of Eq. (41) is modified due to tube compression, which enhances the maximum tube confinement strength. One thus obtains the self-consistent equation: 
\begin{equation}
{{\gamma }_{y,abs}}{{\left( \frac{1+0.5\gamma _{y}^{2}}{\sqrt{1+0.25\gamma _{y}^{2}}} \right)}^{3}}=\lambda {{\gamma }_{y}}
\end{equation}
where $\lambda$ is the stress partition factor of Eq. (31). Note that $_e$Ge and the equilibrium tube diameter factors cancel out as they did in Eq. (41), and ${{\gamma }_{y,abs}}=0.5\ (1.5)$ for $\kappa =1(2.5).$

For all physically sensible values of $\lambda = 0.5-1$,  and both choices of $\kappa$, we find that Eq. (50) has no solution. In essence, we predict that on time scales prior to chain retraction the entanglement network is unbreakable, as in a crosslinked rubber. This result may be related to the suggestion of Wang et. al. that in fast extension or shear the entanglement network ``locks upÕÕ. Moreover, it suggests that chain retraction is a prerequisite to initiating irreversible motions (e.g., flow, loss of elastic recovery, stress overshoot) and the attendant tube softening and microscopic yielding per our analysis of the rigid PP chain model. In this sense, chain retraction is an elementary disentanglement process required for a transition from elastic to viscous behavior, a viewpoint generically argued for by Wang et. al. [17-19,22]. However, our present analysis does not address the issue of the time scale at which chain retraction occurs.

How robust is the above conclusion to our theoretical and numerical assumptions? For instance, the analysis can be repeated assuming tube dilation does not modify either Ge nor the cross sectional area, A. The latter assumption does not seem physical to us, but we pursue it in order to illustrate the sensitivity of our predictions for microscopic yielding to microscopic assumptions. This scenario corresponds to modifying Eq. (50) as:
\begin{equation}
{{\gamma }_{y,abs}}\left( \frac{1+0.5\gamma _{y}^{2}}{\sqrt{1+0.25\gamma _{y}^{2}}} \right)=\lambda {{\gamma }_{y}}
\end{equation}
For $\kappa = 2.5$, Eq. (51) again predicts that the tube does not break. For $\kappa = 1$, microscopic yielding is predicted for essentially all values of $\lambda$  in the range of $0.5 Ð 1$. These differing conclusions reinforce the qualitative sensitivity of the answer to the question of whether the tube breaks to the quantitative value of the maximum tube confinement force. We emphasize that since $\kappa = 2.5$ leads to very good agreement between theory and simulations for the full anharmonic PP distribution function [28,36], we believe the conclusion based on either Eq. (50) or Eq. (51) that there is \emph{no} microscopic absolute yielding  if chains remain fully stretched per a rubber network is the most reliable one.  

Another alternative calculation, which is again in conflict with our rubber-like network model for unrelaxed stretched chains, is to assume $G_e$ \emph{does} scale as the inverse power of the dilated tube diameter squared (hence, the plateau modulus grows with deformation, a type of hardening due to tube compression), and the cross-sectional area A scales as the compressed tube diameter squared. These two changes in the microscopic yielding condition exactly cancel, and one recovers Eq. (51). Finally, yet another possibility is that $G_e$ hardens as the tube diameter is compressed, but $A$ is invariant to deformation; this ansatz seems very unphysical to us.

Thus, the concept of an unbreakable tube on time scales short relative to stretch relaxation has a significant degree of robustness. Addressing the temporal aspects of the crossover requires a force-level treatment of the mechanical stability of stretched chains in entangled liquids. This is the critical and controversial issue of the existence of a ``grip forceÕÕ and an ``entropic barrierÕÕ to retraction discussed by Wang et.al. [17-22]. This topic will be addressed in paper III of this series.

\section{Summary}
We have employed a first-principles, force-level statistical mechanical approach to self-consistently construct the anharmonic tube confinement field and its behavior under nonlinear deformations for entangled fluids composed of infinitely thin needles and of disconnected-PP-level chains on time scales both long and short relative to contour-length relaxation. For the former two systems, deformation-induced orientation leads to tube dilation. In the absence of chain stretch relaxation we find a compression of the tube. 

	Knowledge of how the confinement field changes with polymer transverse displacement allows the calculation of a maximum transverse force that keeps a polymer localized in a tube. The condition for when this entanglement force can be overcome by an external force associated with macroscopic deformation in the absence of irreversible dynamical processes (nonlinear elastic limit) defines the concept of ``microscopic absolute yielding.ÕÕ For needles and contour-length relaxed d-PP chains, a force imbalance is predicted to occur at a stress of order the equilibrium plateau shear modulus. However, for unrelaxed stretched chains, tube compression increases the mechanical strength of the tube, stabilizes transverse confinement, and under almost all physically-sensible scenarios we can think of no force imbalance is predicted. This suggests that the crossover from elastic to irreversible viscous response is likely linked to the chain retraction process. 

The above results correspond to different physics than in existing phenomenological models (e.g., DE [2,3], GLaMM [7]) that are based on the classic idea of an unbreakable and non-deformable tube (other than deformations that follow the macroscopic affine strain). The idea that the entanglement network is fragile, and can be destroyed at strains of order unity or a stress of order the linear modulus, is qualitatively consistent in a general sense with the interpretation of macroscopic experiments and phenomenological arguments of Wang et. al.[17-22]. In particular, for a stress controlled creep experiment, or an abrupt step strain experiment, we do predict absolute microscopic yield can occur, if (or after) chain stretch is relaxed. But the full problem is subtle if chains are stretched precisely because of the ambiguity surrounding the time scale and dynamics of contour length relaxation. Thus, this paper cannot draw definitive conclusions about laboratory continuous startup shear or extension experiments for flexible chain melts or solutions. The nonlinear elastic analysis is also a limiting scenario, and involves subtleties associated with numerical prefactors of order unity that can have major implications. Moreover, the consequences of finite rate deformation and relaxation processes that occur in parallel with nonlinear elastic effects requires a full dynamical treatment. 

 Our results suggests new experiments to test the proposed non-classical ideas, especially step strain and creep measurements on highly entangled rod-like polymers (e.g., microtubules). The latter is being pursued by Robertson-Anderson and coworkers using F-actin [68,69] and DNA [70]. Even more definitive tests of our theory may be performed via simulations of entangled needles under various nonlinear deformations. In the following paper II [71] we present a full dynamical analysis of startup continuous shear for needles and d-PP chain models in the limit of full contour-length relaxation. 

\acknowledgements{KSS thanks Shi-Qing Wang for many stimulating, informative and motivating discussions (and arguments) over the years. The later stages of this work was supported by DOE-BES via the Frederick Seitz Materials Research Laboratory Grant  DE-FG02-07ER46471 (KSS) and by the Advanced Materials Fellowship of the American Philosophical Society (DMS). }


\begin{thebibliography}{10}

\bibitem{ref1} P. G. de Gennes, Scaling Concepts in Polymer Physics, Cornell University Press: Ithaca NY (1979).
\bibitem{ref2} M. Doi and S. F. Edwards, The Theory of Polymer Dynamics, (Oxford University Press, Oxford, 1986).
\bibitem{ref3} M.Doi and S. F. Edwards, J.Chem.Soc. Faraday II, 74, 1789, 1802, 1818 (1978); 75, 38 (1979).
\bibitem{ref4} P. G. de Gennes, J. Chem. Phys. 55 572 (1971).
\bibitem{ref5} T. C. B. McLeish, Adv. Phys. 51, 13709 (2002).
\bibitem{ref6} L. J. Fetters, D. J. Lohse, D. Richter, T. A. Witten, and A. Zirkel, Macromolecules, 27, 4639 (1994).
\bibitem{ref7} R. S. Graham, A. E. Likhtman, T. C. B. McLeish, and S. T. Milner, J. Rheol. 47, 1171 (2003).
\bibitem{ref8} G. Marrucci, J. Non-Newtonian Fluid Mech., 62, 279 (1996).
\bibitem{ref9} W. W. Graessley, Adv.Poly.Sci., 47, 67 (1982).
\bibitem{ref10} G. Marrucci and G. Ianniruberto, J. Non-Newtonian Fluid Mech. 82, 275 (1999).
\bibitem{ref11} G. Ianniruberto and G. Marrucci, J. Non-Newtonian Fluid Mech., 95, 363 (2000).
\bibitem{ref12} H. J. Unidad and G. Ianniruberto, Rheo.Acta, 53, 191 (2014).
\bibitem{ref13} G. Ianniruberto and G. Marrucci, J.Rheol., 58, 89 (2014).
\bibitem{ref14} F. Snijkers, R. Pasquino, P. D. Olmsted, D. Vlassopoulos, J. Phys. Condensed Matter 27, 473002(2015)
\bibitem{ref15} A. E. Likhtman, J. Non-Newtonian Fluid Mech. 157, 158 (2009).
\bibitem{ref16} R. G. Larson, J. Poly. Sci. B: Poly. Phys. 45, 3240 (2007). 
\bibitem{ref17} S. Q. Wang, Soft Matter, 11, 1454 (2015); J. Poly. Sci. Poly. Phys. 46, 2660 (2008). 
\bibitem{ref18} S. Q. Wang, Y. Wang, S. Cheng, X. Li, X. Zhu, and H. Sun, Macromolecules, 46, 3147 (2013).
\bibitem{ref19} S. Q. Wang, S. Ravindranath, Y. Wang, and P. Boukany, J. Chem. Phys. 127, 064903 (2007).
\bibitem{ref20} S. Q. Wang, S. Ravindranath, and P. E. Boukany, Macromolecules 44, 183 (2011).
\bibitem{ref21} Y. Wang and S. Q. Wang, J. Rheol. 53, 1389 (2009).
\bibitem{ref22} P. E. Boukany, S. Q. Wang, and X. Wang, Macromolecules 42, 6261 (2009).
\bibitem{ref23} R. S. Graham, E. P. Henry, and P. D. Olmsted, Macromolecules, 46, 9849 (2013)
\bibitem{ref24} Y. Lu, L. An, S. Q. Wang, Z.-G. Wang, Macromolecules 48, 4164 (2015); ACS MacroLett., 2, 561 (2013); 3, 569 (2014).
\bibitem{ref25} J. Cao and A. E. Likhtman , ACS Macro Lett., 4, 1376 (2015).
\bibitem{ref26} D. M. Sussman and K. S. Schweizer, Phys. Rev. E, 83, 061501 (2011).  
\bibitem{ref27} D. M. Sussman and K. S. Schweizer, Phys. Rev. Lett., 107, 078102 (2011).
\bibitem{ref28} D. M. Sussman and K. S. Schweizer, Phys. Rev. Lett., 109, 168306 (2012).
\bibitem{ref29} D. M. Sussman and K. S. Schweizer, J. Chem. Phys., 139, 234904 (2013).
\bibitem{ref30} D. M. Sussman, W. S. Tung, K. I. Winey, K. S. Schweizer and R. A. Riggleman, Macromolecules, 47, 6462 (2014).
\bibitem{ref31} D. M. Sussman and K. S. Schweizer, J.Chem. Phys., 135, 131104 (2011).  
\bibitem{ref32} D. M. Sussman and K. S. Schweizer, Macromolecules, 45, 3270 (2012).
\bibitem{ref33} D. M. Sussman and K. S. Schweizer, Macromolecules, 46, 5684 (2013).
\bibitem{ref34} E. Ben-Naim, G. S. Grest, T. A. Witten and A. R. C. Baljon, Phys.Rev.E 53, 1816 (1996).
\bibitem{ref35} A. E. Likhtman and M. Ponmurigan, Macromolecules, 47, 1470 (2014); A. E. Likhtman, Soft Matter, 10, 1895 (2014).
\bibitem{ref36} S. D. Anogiannakis, C. Tzoumanekas, and D. N. Theodorou, Macromolecules 45, 9475 (2012); C. Tzoumanekas and D. N. Theodorou, Macromolecules 39, 4592 (2006).
\bibitem{ref37} J. Cai, J. Qin and S. T. Milner, Macromolecules, 48, 99 (2015); J. Qin and S. T. Milner, Macromolecules, 47, 6077 (2014).
\bibitem{ref38} C. Baig, V. G. Mavrantzas and M. Kroger, Macromolecules 43, 6886 (2010).
\bibitem{ref39} G. Ronca, J. Chem. Phys., 79, 1031 (1983).
\bibitem{ref40} S. F. Edwards and J. W. V. Grant, J. Phys. A, 6, 1169 (1973).
\bibitem{ref41} K. S. Schweizer, J. Chem. Phys., 91, 5822 (1989).
\bibitem{ref42} K. S. Schweizer, M. Fuchs, G. Szamel, M. Guenza and H. Tang, Macromolecular Theory and Simulation, 6, 1037(1997)
\bibitem{ref43} R. Kimmich and N. Fatkullin , Adv. Polym. Sci., 170, 1 (2004).
\bibitem{ref44} N. Fatkullin and R. Kimmich, Macromol. Symp. 146, 103, (1999).
\bibitem{ref45} W. Hess, Macromolecules, 21, 2620 (1988).
\bibitem{ref46} M. Fixman, J. Chem. Phys., 89, 3892 and 3912 (1988).
\bibitem{ref47} K. Kawasaki, Mod. Phys. Lett. B, 4, 913 (1990).
\bibitem{ref48} M. Guenza, J. Chem. Phys., 110, 7574 (1999).
\bibitem{ref49} M. Guenza, Phys. Rev. E, 89, 052603 (2014).
\bibitem{ref50} K. S. Kim, S. Dutta and Y. Jho, Soft Matter 11, 7932 (2015).
\bibitem{ref51} G. Szamel, Phys. Rev. Lett. 70, 3744 (1993); G. Szamel and K. S. Schweizer, J. Chem. Phys., 100, 3127 (1994).
\bibitem{ref52} A. S. Lodge, Rheo. Acta 28, 351 (1989).
\bibitem{ref53} K. S. Schweizer and E. J. Saltzman, J. Chem. Phys., 118, 1181 (2003); K. S. Schweizer, J. Chem. Phys.,123,244501(2005); K. Chen. E. J. Saltzman and K. S. Schweizer, Ann. Rev. Condensed Matter Physics, 1, 277 (2010).
\bibitem{ref54} B. Wang, et. al.,  Phys. Rev. Lett. 104, 118301 (2010).
\bibitem{ref55} R. M. Anderson and D. E. Smith, Phys. Rev. Lett., 99, 126001 (2007).
\bibitem{ref56} V. Kobelev and K. S. Schweizer, Phys. Rev. E 71, 021401 (2005); J. Chem. Phys. 123, 164903 (2005).
\bibitem{ref57} S. Ramanathan and D. C. Morse, Phys. Rev. E 76 010501(R) (2007). 
\bibitem{ref58} C. P. Broedersz and F. C. MacKintosh, Rev. Mod. Phys., 86, 995 (2014).
\bibitem{ref59} A. E. Likhtman, Macromolecules 38, 6128 (2005).
\bibitem{ref60} D. M. Nair and J. D. Schieber, Macromolecules 39, 3386 (2006); R. N. Khaliullin and J. D. Schieber, Phys. Rev. Lett. 100, 188302 (2008).
\bibitem{ref61} H. Eyring, J. Chem. Phys., 4, 283 (1938). 
\bibitem{ref62} R. Everaers, S. K. Sukumaran, G. S. Grest, C. Svaneborg, A. Sivasubramanian, and K. Kremer, Science 303, 823 (2004).
\bibitem{ref63} T. A. Kavassalis and J. Noolandi, Phys. Rev. Lett. 59, 2674 (1987).
\bibitem{ref64} R. Everaers, Phys. Rev. E 86, 022801 (2012).
\bibitem{ref65} D. Auhl, J. Ramirez, A. E. Likhtman, P. Chambon and C. Fernyhough, J. Rheol. 52, 801 (2008). 
\bibitem{ref66} Y. Wang and S. Q. Wang, Macromolecules, 52, 1275 (2008); G. Liu, H. Sun and S. Q. Wang, Macromolecules 57, 89 (2013).
\bibitem{ref67} P. G. deGennes, MRS Bulletin, January 1991, p.2.
\bibitem{ref68} T. T. Falzone and R. M. Robertson-Anderson, ACS MacroLett. 4, 1194 (2015).
\bibitem{ref69} T. T. Falzone, S. Blair and R. M. Robertson-Anderson, Soft Matter 11, 4418 (2015).
\bibitem{ref70} C. D. Chapman and R. M. Robertson-Anderson, Phys. Rev. Lett., 113, 098303(2014).
\bibitem{ref71} K. S. Schweizer and D. M. Sussman, following paper II. 

\end{thebibliography}
\end{document}